\documentclass[pdflatex,sn-basic,Numbered]{sn-jnl}

\usepackage{graphicx}%
\usepackage{multirow}%
\usepackage{amsmath,amssymb,amsfonts}%
\usepackage{amsthm}%
\usepackage{mathrsfs}%
\usepackage[title]{appendix}%
\usepackage{xcolor}%
\usepackage{textcomp}%
\usepackage{manyfoot}%
\usepackage{booktabs}%
\usepackage{algorithm}%
\usepackage{algorithmicx}%
\usepackage{algpseudocode}%
\usepackage{listings}%

\usepackage[nobreak=true]{mdframed}
\usepackage{subcaption}
\usepackage{tabularray}
\usepackage{hyperref}
\usepackage{thm-restate}

\theoremstyle{thmstyleone}

\newtheorem{proposition}{Proposition} 
\newtheorem{lemma}{Lemma}

\theoremstyle{thmstyletwo}%
\newtheorem{corollary}{Corollary}%
\newtheorem{remark}{Remark}%

\newcommand{\AlgInput}[1]{\textbf{Input:} #1}
\newcommand{\AlgOutput}[1]{\textbf{Output:} #1}

\newcommand{\AlgDS}[1]{\textbf{Data Structure:} #1}
\newcommand{\AlgParameter}[1]{\textbf{Parameter:} #1}

\DeclareMathOperator*{\argmin}{arg\,min}

\newcommand{\problem}[3]{
	\begin{mdframed}
		\textsc{#1}\\
		~
		\begin{tabular}{rp{0.8\columnwidth}}
			\textbf{Instance}: & #2\\
			\textbf{Question}: & #3
		\end{tabular}
	\end{mdframed}
}

\newcommand{\pproblem}[4]{
	\begin{mdframed}
		\textsc{#1}\\
		~
		\begin{tabular}{rp{0.8\columnwidth}}
			\textbf{Instance}: & #2\\
			\textbf{Parameters}: & #4\\
			\textbf{Question}: & #3
		\end{tabular}
	\end{mdframed}
}

\newcommand{\msap}{\textsc{MSA-DH} problem}

\newcommand{\mhsp}{\textsc{MHS} problem}

\newcommand{\sa}{Steiner arborescence}
\newcommand{\sn}{Steiner nodes}
\newcommand{\instance}[1]{\langle #1 \rangle}

\newcommand{\Qm}{\vec{Q}_{\dimension}}
\newcommand{\level}[1]{\mathtt{level}(#1)}
\newcommand{\zerom}{0^{\dimension}}

\newcommand{\cost}[1]{\mathtt{COST}(#1)}
\newcommand{\optarb}{T_{\mathrm{opt}}}
\newcommand{\solarb}{T}
\newcommand{\opt}{OPT}
\newcommand{\sol}{SOL}
\newcommand{\vertexset}[1]{V(#1)}
\newcommand{\edgeset}[1]{E(#1)}

\newcommand{\solStNode}{\mathtt{S}}

\newcommand{\terminalset}{R}
\newcommand{\modr}{|R|}
\newcommand{\dimension}{m}
\newcommand{\popt}{p_{\mathrm{opt}}}
\newcommand{\qopt}{q_{\mathrm{opt}}}
\newcommand{\p}{p_{\solarb}}
\newcommand{\q}{q_{\solarb}}
\newcommand{\mset}{[\dimension]}

\newcommand{\maxlevel}{\ell_{\mathrm{max}}}

\newcommand{\tpath}[1]{P(#1)}
\newcommand{\hdist}[1]{d(#1)}

\newcommand{\anc}[1]{\mathtt{Ancestor}(#1)}
\newcommand{\des}[1]{\mathtt{Descendant}(#1)}
\newcommand{\parent}[1]{\mathtt{parent}(#1)}
\newcommand{\child}[1]{\mathtt{child}(#1)}

\newcommand{\lca}[1]{\mathtt{LCA}(#1)}

\newcommand{\good}[1]{\mathtt{GOOD}(#1)}
\newcommand{\bad}[1]{\mathtt{BAD}(#1)}
\newcommand{\cg}[1]{\mathtt{CG}(#1)}

\newcommand{\minhitset}[1]{\mathtt{MHS}(#1)}

\newcommand{\mvc}{\mathtt{MVC}}
\newcommand{\vc}{\mathtt{VC}}
\newcommand{\mvcs}{\tau}

\newcommand{\fpt}{\textsf{FPT}}
\newcommand{\pfpt}[1]{\fpt$[#1]$}
\newcommand{\whard}[1]{\textsf{W[#1]}-hard}
\newcommand{\poly}{\textsf{P}}
\newcommand{\np}{\textsf{NP}}

\newcommand{\ztime}{\textsf{ZTIME}}
\newcommand{\dtime}{\textsf{DTIME}}

\newcommand{\Oh}[1]{\mathcal{O}\left(#1\right)}
\newcommand{\Ohs}[1]{\tilde{\mathcal{O}}(#1)}

\newcommand{\univ}{U}
\newcommand{\setsys}{\mathcal{F}}

\begin{document}

\title[MSA-DH Problem]{Parameterized Algorithms for the Steiner Arborescence Problem on a Hypercube}

\author*[1,2,3]{\fnm{Sugyani} \sur{Mahapatra}}\email{msugyani@cse.iitm.ac.in}

\author*[1,2,3]{\fnm{Manikandan} \sur{Narayanan}}\email{nmanik@cse.iitm.ac.in}

\author*[1]{\fnm{N. S. } \sur{Narayanaswamy}}\email{swamy@cse.iitm.ac.in}

\affil[1]{\orgdiv{Department of Computer Science and Engineering}, \orgname{Indian Institute of Technology (IIT) Madras}, \orgaddress{\city{Chennai}, \state{Tamil Nadu}, \country{India}}}

\affil[2]{\orgdiv{Center for Integrative Biology and Systems Medicine}, \orgname{IIT Madras}, \orgaddress{\city{Chennai}, \state{Tamil Nadu}, \country{India}}}

\affil[3]{\orgdiv{Robert Bosch Centre for Data Science and Artificial Intelligence}, \orgname{IIT Madras}, \orgaddress{\city{Chennai}, \state{Tamil Nadu}, \country{India}}}

\abstract{
    Motivated by a phylogeny reconstruction problem in evolutionary biology, we study the minimum Steiner arborescence problem on directed hypercubes (\textsc{MSA-DH}).
    Given $m$, representing the directed hypercube $\vec{Q}_m$, and a set of terminals $\terminalset$, the problem asks to find a Steiner arborescence that spans $\terminalset$ with minimum cost. As $\dimension$ implicitly represents $\Qm$ comprising $2^{\dimension}$ vertices, the running time analyses of traditional Steiner tree algorithms on general graphs does not give a clear understanding of the actual complexity of this problem.
	We present algorithms that exploit the structure of the hypercube and run in time polynomial in $\modr$ and $\dimension$.

	We explore the \msap ~on three natural parameters -- $\modr$, and two above-guarantee parameters, number of \sn~$p$ and penalty $q$. For above-guarantee parameters, the parameterized \msap ~take $p \geq 0$ or $q\geq 0$ as input, and outputs a \sa ~with at most $\modr + p - 1$ or $\dimension + q$ edges respectively.
	We present the following results ($\tilde{\mathcal{O}}$ hides the polynomial factors):
	\begin{enumerate}
		\item An exact algorithm that runs in $\Ohs{3^{\modr}}$ time.
		\item A randomized algorithm that runs in $\Ohs{9^q}$ time with success probability $\geq 4^{-q}$.
        \item An exact algorithm that runs in $\Ohs{36^q}$ time.
		\item A $(1+q)$-approximation algorithm that runs in $\Ohs{1.25284^q}$ time.
		\item An $\Oh{p\maxlevel}$-additive approximation algorithm that runs in $\Ohs{\maxlevel^{p+2}}$ time, where $\maxlevel$ is the maximum distance of any terminal from the root.
	\end{enumerate}
}

\maketitle

\section{Introduction}

\subsection{Problem and Motivation}
A long-standing problem inspired by Darwin's theory of evolution is to reconstruct the ``tree of life'' or phylogeny relating all current-day species from their DNA/protein sequences. The problem is to specifically infer a phylogenetic tree~\cite{phylogenyBook} that relates the observed sequences of current-day species via the unobserved sequences of their common ancestral species. We can model this problem as the hypercube variant of the Steiner tree problem, by considering the binary representation of the observed sequences as input terminals and unobserved ancestral sequences as Steiner nodes. Finding a Steiner tree in an undirected hypercube solves the phylogeny problem when mutations are reversible (i.e., both $0\rightarrow1$ and $1\rightarrow0$ bit flips/changes are allowed), and a directed hypercube solves the problem when mutations are irreversible (i.e., only $0\rightarrow1$ bit flips are possible). The former is better studied, and the focus of this work is on open questions related to the latter problem of finding a Steiner Arborescence in a directed hypercube. Given that the Steiner tree problem is one of Karp's 21 \np-complete problems~\cite{intractableProblems}, there is rich work on the parameterized complexity of the Steiner tree problem on general as well as special-case graphs. For instance in general graphs, the Steiner tree problem is \fpt~on the number of input terminals $\modr$ as parameter~\cite{Dreyfus-Wagner, steinerSurvey}, but \whard{2} on the number of Steiner nodes $p$ as parameter~\cite{St-PAAlgo-p}. Despite this rich literature, the directed hypercube variant of the Steiner tree focused in this study has not been addressed before.  

We now formally state our problem.
For an $\dimension \geq 1$, the $\dimension$-hypercube $Q_{\dimension}$ has all the $2^{\dimension}$ binary strings of length $\dimension$ as the vertex set, i.e., $\vertexset{Q_{\dimension}} = \{0,1\}^{\dimension}$, and there exists an edge between two vertices $t_1$ and $t_2$ if they differ exactly at a single bit position.
Every bit position is also known as a `\emph{character}'~\cite{WP-FPT-journal, WP-FPT-conf}.
$\Qm$ denotes the directed hypercube rooted at $\zerom$ such that the edges are oriented away from the root towards vertices with a higher number of 1s. 
For an arborescence $\solarb$, $\cost{\solarb}$ denotes $ |\edgeset{\solarb}|$. Then \msap ~is stated as follows.

\problem{\sloppy Minimum Steiner Arborescence on Directed Hypercube (MSA-DH) Problem}{
	\sloppy An integer $\dimension$, and a set of terminals $\terminalset \subseteq \vertexset{\Qm}$. 
}{
	\sloppy Find a \sa~$\solarb$ in $\Qm$, rooted at $\zerom$, that spans $\terminalset$ with the minimum cost. 
}
As $\zerom$ must be in any output \sa, we assume without loss of generality that $\zerom \in \terminalset$.
Note that in a solution arborescence $\solarb$, the path from the root to a terminal encodes a sequence of bit flips/changes, one at a time, to transform or mutate the root to become the terminal.  
Thus, each edge in $\solarb$ is associated with the bit flip/change of one of the $\dimension$ characters.
This problem models a variant of the phylogeny problem as discussed above, called the Camin-Sokal Phylogeny~\cite{CSP}, and is also motivated by a related model in network science known as the \emph{Expander Hierarchy Model}~\cite{expanderHierarchy}, which uses a hierarchical tree representation of graph objects to efficiently compute properties of a dynamic graph.

We cannot hope for an algorithm that runs in time polynomial in $\modr$ and $\dimension$, since the \msap~on hypercubes encode a generalized version of the minimum vertex cover problem and thus, is \np-complete for both undirected~\cite{WP-NPC} and directed~\cite{CSP-NPC} cases. 
Owing to the exponential size of the underlying graph, existing algorithms for Steiner trees on general graphs will run in time exponential in the input size.
There are some parameterized complexity ~\cite{WP-FPT-conf,WP-FPT-journal,lam2010imperfect} and approximability~\cite{WP-APXC} results on the penalty parameter $q$ (defined as the extra cost incurred by the solution on the characters) for the undirected version of the problem. 
However, in the directed case, i.e., the \msap, there are no such results to the best of our knowledge.
The case when the optimal penalty $\qopt = 0$ is called the \emph{Perfect Phylogeny problem} and can be solved in polynomial time~\cite{perfectPhylogeny-conf,perfectPhylogeny-journal,perfectPhylogenyCSP,phylogenyBasics}.
A summary of the literature is in \autoref{tab:lit-survey} and provides the context for our work.

\begin{table}[h]
	\centering
	\resizebox{\textwidth}{!}{
		\begin{talltblr}[
			caption = {Brief review of literature on Steiner trees in different types of graphs. The black and purple colors show existing literature and our contributions respectively.},
			label = {tab:lit-survey},
			note{1} = {First algorithm}, note{2} = {Current best known algorithm}, note{3} = {$\tilde{\mathcal{O}}$ notation hides the running time factors that are polynomial in $\modr$ and $\dimension$}, note{4} = {PAS is Parameterized Approximation Scheme, EPAS is Efficient Parameterized Approximation Scheme, PSAKS is Polynomial-Size Approximate Kernelization Scheme; (Also, we use standard abbreviations FPT for Fixed-Parameter Tractable, and DAG for Directed Acyclic Graph)}
			]{
				|| Q[l,m] || Q[c,m] || Q[l,m] | Q[l,m] | Q[l,m] | Q[l,m] | Q[l,m] ||
			}
			\hline \hline
			\SetCell[r=2]{c} & \SetCell[r=2]{c} Parameter & \SetCell[c=2]{c,m} General Graphs & & \SetCell[r=2]{m,c} DAG & \SetCell[c=2]{c,m} Hypercube &\\			
			\hline
			&  & \SetCell{c} Undirected & \SetCell{c} Directed &  & \SetCell{c} Undirected & \SetCell{c} Directed \\
			
			\hline \hline			
			\SetCell[r=3]{c} \fpt~Algorithm 
			& $\modr$ 
			& \SetCell[c=1]{l} {$\Ohs{3^{\modr}}$ -- (Dreyfus-\\Wagner)~\cite{Dreyfus-Wagner,FPTAlgoBook}~\TblrNote{1} \\ $\Ohs{2^{\modr}}$~\cite{steinerSurvey}~\TblrNote{2}} 
			& {$\Ohs{3^{\modr}}$ -- (Dreyfus-\\Wagner)~\cite{Dreyfus-Wagner,FPTAlgoBook}~\TblrNote{1} \\ $\Ohs{2^{\modr}}$~\cite{steinerSurvey}~\TblrNote{2}}
			& Hsu et.al.~\cite{steinerTreeDAG} 
			& \SetCell[r=2]{c,m} OPEN
			& {\color{purple} $\Ohs{3^{\modr}}$  cf. [\autoref{thm:r:fpt}]} \\
			
			\hline			
			& $p$ 
			& \SetCell[c=1]{l} \whard{2}~\cite{St-PAAlgo-p} 
			& \whard{2}~\cite{St-PAAlgo-p}
			& \SetCell[c=1]{c} OPEN
			&& \SetCell[c=1]{c} OPEN \\ 
			
			\hline			
			& $q$ 
			& \SetCell[c=1]{c} NA
			& \SetCell[c=1]{c} NA
			& \SetCell[c=1]{c} NA
			& $\Ohs{21^q}$~\cite{WP-FPT-conf,WP-FPT-journal}~\TblrNote{1}
			& \SetCell{l,m} {\color{purple} $\Ohs{9^q}$ with $\Pr \geq 4^{-q}$ \\ cf. [\autoref{thm:q:fpt}] \\ $\Ohs{36^q}$ cf. [\autoref{lemma:q:fpt:new}]} \\
			
			\hline \hline
			{Approximation \\ Algorithms} 
			& -- 
			& 2-approx~\cite{approxAlgoBook} 
			& {No $\Ohs{\log^{2-\epsilon} \modr}$ \\ unless \np~$\subseteq$~\ztime \\ \cite{steinerSurvey,St-PAAlgo-p,ST-approx}. \\ No $\Ohs{(1-\epsilon) \log \modr}$ \\ unless \poly~$=$~\np~\cite{St-PAAlgo-p}} 
			& {$\modr^\epsilon$ exists but \\ no $\frac{\ln \modr}{4}$ unless \\ \np~$\subseteq$~\dtime~\cite{St-DAG-approximation}} 
			& {$\Oh{q^2}$-additive \\ approximation~\cite{WP-approx} \\ 2-approximation \\ (Kou et al.)~\cite{approxAlgoBook}} 
			& {\color{purple} $\Oh{\qopt}$-approximation \\ cf. [\autoref{remark:q:approx}]; \\ $\Ohs{\popt \maxlevel^2}$-additive \\ approximation \\ cf. [\autoref{remark:p:approx}]}
			\\
			
			\hline \hline
			\SetCell[r=3]{l} {Parameterized \\ Approximation \\ Algorithm} 
			& $\modr$ 
			& { PAS and \\ PSAKS; no \\ polynomial kernel \\ unless \np ~\\$\subseteq$ co-\np \cite{steinerSurvey}. }
			& \SetCell[r=1,c=1]{c} OPEN
			& \SetCell[r=1,c=1]{c} OPEN
			& \SetCell[r=3,c=1]{c} OPEN
			& \SetCell[r=1,c=1]{c} OPEN \\
			
			\hline
			& $p$
			& {EPAS ($\epsilon > 0$) and \\ PSAKS~\cite{St-PAAlgo-p}}
			& {EPAS exists but \\ no PSAKS. \textsf{W[1]}\\-hard for $f(p)$-\\ approximation~\cite{St-PAAlgo-p}.}
			& \SetCell{c} EPAS~\cite{St-PAAlgo-p}
			&& {\color{purple} $\Oh{p\maxlevel}$-additive \\ approximation in \\ $\Ohs{\maxlevel^{p+2}}$ cf. [\autoref{thm:p:approx}]}\\
			
			\hline
			& $q$
			& \SetCell[r=1,c=1]{c} NA
			& \SetCell[r=1,c=1]{c} NA
			& \SetCell[r=1,c=1]{c} NA
			&& {\color{purple} $(1+q)$-approximation \\ in $\Ohs{1.25284^q}$ \\ cf. [\autoref{thm:q:approx}]} \\
			
			\hline \hline 
			
		\end{talltblr}
	}
\end{table}

\subsection{Our Results}
We consider three parameters -  number of terminals $\modr$, number of \sn~$p$ and penalty $q$. To consider the parameterizations with respect to $p$ and $q$ as above guarantee parameters, we observe the following lower bounds on the cost of any solution.  Since $\modr$ is the number of terminals, any solution has at least $\modr - 1$ edges. Thus the cost of a \sa~$\solarb$~is given by $\cost{\solarb} = \modr - 1 + \p$ where $\p$ is the number of \sn.
For the parameter $q$, note that a character with the same bit value across all the terminals does not play a significant role in the construction of the \sa~(cf. \autoref{lemma:normalization}); thus we assume that for each of the $\dimension$ characters, the terminal set can be partitioned into two non-empty sets -- those in which the character is 0 and those in which it is 1. Thus in any solution, the value of each character must change at least once along some edge in $\solarb$. 
Thus, $\cost{\solarb} \geq \dimension$, and the penalty $\q$ of the arborescence $\solarb$ is given as $\q = \cost{\solarb} - \dimension$~\cite{WP-FPT-journal,WP-FPT-conf}.
The optimum number of \sn~is denoted by $\popt = \min\limits_{\solarb} \p$; and the optimal penalty is denoted by $\qopt = \min\limits_{\solarb} \q$.

The parameterized (decision) version of the problem is now defined as:
\pproblem{\sloppy Parameterized \msap}{
	\sloppy An integer $\dimension$, and a set of terminals $\terminalset \subseteq \vertexset{\Qm}$. 
}{
	\sloppy Is there a \sa~$\solarb$ in $\Qm$, rooted at $\zerom$, that spans $\terminalset$ and has cost at most $\dimension + q$ or at most $\modr + p - 1$? 
}{
	\sloppy $\modr$, penalty $q \ge 0$, number of \sn~$p \ge 0$.
}
In the decision algorithms we attempt to compute the optimal \sa, thus giving the value of $\popt$ and $\qopt$. Thus, if $q \ge \qopt$ ($p \ge \popt$), we output the \sa~$\solarb$; otherwise when $q < \qopt$ ($p < \popt$), the algorithms output ``No''. 
We use the notation \pfpt{k} to denote an \fpt~algorithm parameterized by $k$.

On the contrary, with parameterized approximation algorithms, we attempt to compute an approximate solution with better running times.
We adopt the definition used by Dvor{\'a}k et al.~\cite{St-PAAlgo-p}, i.e., if $q \ge \qopt$ ($p \ge \popt$), we output an approximate solution with performance ratio $\alpha$ on the cost; otherwise the output is arbitrary.

We first present a dynamic programming formulation similar to the \pfpt{\modr}-algorithm for the Steiner tree problem in undirected graphs by Dreyfus and Wagner~\cite{Dreyfus-Wagner}.
\begin{restatable}{theorem}{rfpt}
	\label{thm:r:fpt}
	The \msap~can be solved in $\Oh{3^{\modr} \modr \dimension + \modr^2 \dimension^2}$ time and $\Oh{2^{\modr} \log {\modr\dimension} }$ space.
\end{restatable}

For the parameter $q$, the undirected version of the problem admits a randomized \pfpt{q}-algorithm (see \autoref{tab:lit-survey}). 
Inspired by this algorithm, we present a randomized \pfpt{q}-algorithm for the parameterized \msap~as well.
\begin{restatable}{theorem}{qfpt}
	\label{thm:q:fpt}
	\sloppy The parameterized \msap~can be solved in $\Oh{9^q q^3 + q\modr \dimension^2}$ time with success probability $\geq 4^{-q}$.
\end{restatable}
The algorithm can be derandomized using a depth-bounded search tree in $\Ohs{36^q}$ time (cf. \autoref{lemma:q:fpt:new}).

We next present a novel parameterized approximation algorithm with improved running time but at the expense of exactness by using the exact algorithm for the minimum vertex cover problem~\cite{FPTAlgoBook}.
\begin{restatable}{theorem}{qapprox}
	\label{thm:q:approx}
	There is a $(1+q)$-approximation algorithm for the parameterized \msap ~that runs in $\Ohs{1.25284^q}$ time, where $\tilde{\mathcal{O}}$ hides the polynomial factors in the input size.
\end{restatable}

The Steiner tree problem is infamously \whard{2} on the  parameter $p$, the number of \sn, for general graphs and has not been explored for DAGs and undirected hypercubes.
To the best of our knowledge, the parameterized complexity of the problem on directed hypercubes is yet to be known.
We present a novel parameterized approximation algorithm on the parameters $p$ and $\maxlevel$ for the same, where $\maxlevel$ is the maximum distance of any terminal from the root.
\begin{restatable}{theorem}{papprox}
	\label{thm:p:approx}
	There is a $\Oh{p\maxlevel}$-additive approximation algorithm for the parameterized \msap ~that runs in $\Oh{\maxlevel^{p+2} + \maxlevel^2 \modr^2 \dimension}$ time, where $\maxlevel$ is the maximum distance of any terminal from the root.
\end{restatable}

In our algorithms, the key point is to note the polynomial dependence on the $\dimension$ and $\modr$ (for the parameters $p$ and $q$), and the exponential dependence on the parameter.  
The algorithms have all exploited the nature of the directed hypercube, and we believe that the dependence on the parameters can be further improved.  
In particular, we have not been able to improve the running time of the algorithm parameterized by the number of terminals from $\Ohs{3^{\modr}}$ to $\Ohs{2^{\modr}}$.
We also note that while these algorithms can be used to solve the Steiner problem in an undirected hypercube, the resulting tree may not be optimal (e.g., \autoref{remark:r:fpt}) or adhere to the approximation ratios. 
Conversely, results on the Steiner problem in an undirected hypercube cannot of course be assumed to hold ``as is'' for its directed version (where $1\rightarrow0$ character flips/changes are disallowed). These observations motivate our results above on the complexity of the Steiner problem in a directed hypercube.

\section{Preliminaries}
Given the dimension $\dimension$, let $\Qm$ denote the directed hypercube rooted at $\zerom$. 
Let $\terminalset\subseteq \vertexset{\Qm} = \{0,1\}^{\dimension}$ be the input set of terminals.
We denote an optimal \sa~as $\optarb$ and the output \sa~as $\solarb$ ($\optarb^{\terminalset}$ and $\solarb^{\terminalset}$ denote the optimal and output \sa~on the terminal set $\terminalset$ respectively).
$\vertexset{\solarb}$ and $\edgeset{\solarb}$ denote the node and edge set of the arborescence $\solarb$ respectively.
We use $\opt_{\terminalset}$ and $\sol_{\terminalset}$ to denote cost of the optimal and output arborescence respectively for $\terminalset$. 
When $\terminalset$ is clear from the context, we just write $\opt$ and $\sol$.

Throughout the paper, $I = \instance{\dimension,\terminalset}$ denotes a \msap ~instance. We assume $\modr \geq 2$ without loss of generality (otherwise the solution is trivial).
Let $\mset$ denote the set of all $\dimension$ characters for $I$.
For $t\in \vertexset{\Qm}$ and $u \in \mset$, $t[u]$ denotes the value of the $u$\textsuperscript{th} character in $t$.
Also, we use the notation $[i,j]$ to represent the set $\{i, i+1, \ldots, j\}$.

For  $t_1,t_2 \in \vertexset{\Qm}$, let $\hdist{t_1, t_2}$ represent the Hamming distance between $t_1$ and $t_2$.
The vertices of $\Qm$ are partitioned into $\dimension + 1$ levels based on the Hamming distance of the vertices from $\zerom$.
For any node $t \in \vertexset{\Qm}$, $\level{t} = \hdist{\zerom, t}$ represents the level of $t$ in $\Qm$.
We use $\maxlevel$ to represent the maximum level of the terminals i.e. $\maxlevel = \max_{t\in \terminalset}~\level{t}$.
$\Qm^{\ell}$ is the set of nodes in $\Qm$ at level $\ell$.
\sloppy Given any node $t\in \vertexset{\Qm}$, we use $\anc{t}$, $\des{t}$, $\parent{t}$ and $\child{t}$ to represent the sets of ancestors, descendants, parents and children of $t$ respectively. 

For a subset $S \subseteq \vertexset{\Qm}$, the least common ancestor of $S$ in the hypercube, denoted by $\lca{S}$, is that node in the hypercube whose descendants contain $S$, but no child of $\lca{S}$ contain $S$ as its descendants. The $\lca{S}$ can be computed by applying the bitwise `boolean AND' operation on the elements of $S$.

\begin{lemma}
	\label{lemma:normalization}
	Let $u \in \mset$ such that for all $t_1, t_2 \in \terminalset \setminus \{\zerom\}$, $t_1[u] = t_2[u]$. 
	Let $I'=\instance{\dimension',\terminalset'}$, where $[\dimension'] = \mset \setminus \{u\}$ and $\terminalset' = \{t' \mid \exists~ t\in \terminalset \text{ such that } \forall~ v\in[\dimension']~ t'[v] = t[v] \}$. Then
	\[
	\opt_{\terminalset} = 
	\begin{cases}
		\opt_{\terminalset'} & if~ \forall~ t \in \terminalset \setminus \{\zerom\}~ t[u] = 0\\
		\opt_{\terminalset'} + 1 & if~ \forall~ t \in \terminalset \setminus \{\zerom\}~ t[u] = 1
	\end{cases}
	\]
\end{lemma}
\begin{proof}	
	Observe that an optimum solution for $R$ exists with the following canonical property -- if $t[u]=0$ for all terminals, then the character $u$ does not change along any edge in the solution. On the other hand, if $t[u]=1$ for all terminals excluding $\zerom$, then the character $u$ changes once on the only edge leaving $\zerom$, and does not change along any other edge.  This lemma immediately follows from this observation.
\end{proof}
We thus assume every input instance is preprocessed by \autoref{lemma:normalization}, unless mentioned explicitly. As a consequence of this lemma, $\lca{\terminalset} = \zerom$. 
Also note that if  there is no requirement to root the Steiner arborescence at $\zerom$, every optimal \sa ~for $\terminalset$ will be rooted at $\lca{\terminalset}$.

\section{Dynamic Programming on Number of Input Terminals}
Our Dynamic Programming (DP) formulation has similarities to the Dreyfus-Wagner algorithm~\cite{Dreyfus-Wagner}.
The optimal substructure of the DP formulation is based on least common ancestors and is given in \autoref{lemma:optsubstruct}.
It must be noted here that in the recursive subproblems the subsets of the input terminal set are not preprocessed and could contain invariant characters. Thus, the optimal arborescences considered for the subproblems are rooted at \texttt{LCA} of the corresponding subset of the terminal set instead of $\zerom$.

\begin{lemma}
	\label{lemma:optsubstruct}
	For an instance $I$ of the \msap, there exists $\phi \neq \terminalset'\subset \terminalset$ and $\terminalset'' = \terminalset\setminus \terminalset'$ such that $\optarb^{\terminalset}$ can be decomposed into two subtrees that are each optimal \sa s for terminal sets $\terminalset'$ and $\terminalset''$ respectively. 
	As a result,
	\begin{equation}
		\opt_{\terminalset} = \opt_{\terminalset'} + \opt_{\terminalset''} + \hdist{\lca{\terminalset}, \lca{\terminalset'}} + \hdist{\lca{\terminalset}, \lca{\terminalset''}}
	\end{equation}
\end{lemma}

\begin{proof}
	We know that $\lca{\terminalset}$ will be the root of $\optarb^{\terminalset}$. 
	If $\lca{\terminalset}$ has at least two children in $\optarb^{\terminalset}$, then let $x$ be any one of its child.
	Let $\terminalset'$ be the set of all terminals spanned by the subtree of $\optarb^{\terminalset}$ rooted at $x$, denoted by $\solarb_{x}$, and let $\terminalset'' = \terminalset\setminus \terminalset'$. 
	Since $\optarb^{\terminalset}$ is a minimum cost \sa, note that $\terminalset'\neq\phi$. Note also that $\terminalset'' \neq\phi$ (by a similar argument where we let $x$ to be any other child of $\lca{\terminalset}$).  
    So we have a non-trivial partition of $\terminalset$ (see  
	Figure~\ref{fig:dppa-substruct-c2} for an illustration).
	
	\begin{figure}[h]
		\centering
		\begin{subfigure}[b]{0.48\columnwidth}
			\centering
			\includegraphics[scale=0.35]{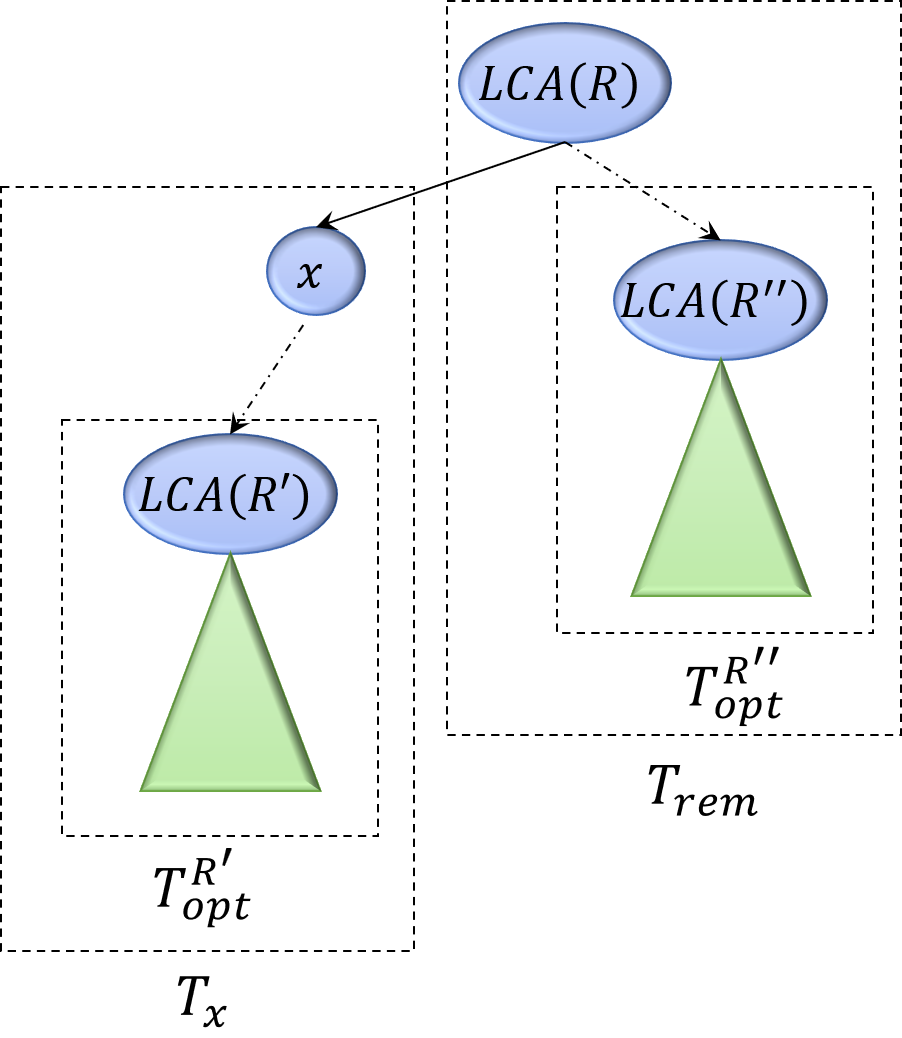}
			\caption{\label{fig:dppa-substruct-c2} When $\lca{\terminalset}$ has at least two subtrees.}
		\end{subfigure}
		\hfil
		\begin{subfigure}[b]{0.45\columnwidth}
			\centering 
			\includegraphics[scale=0.35]{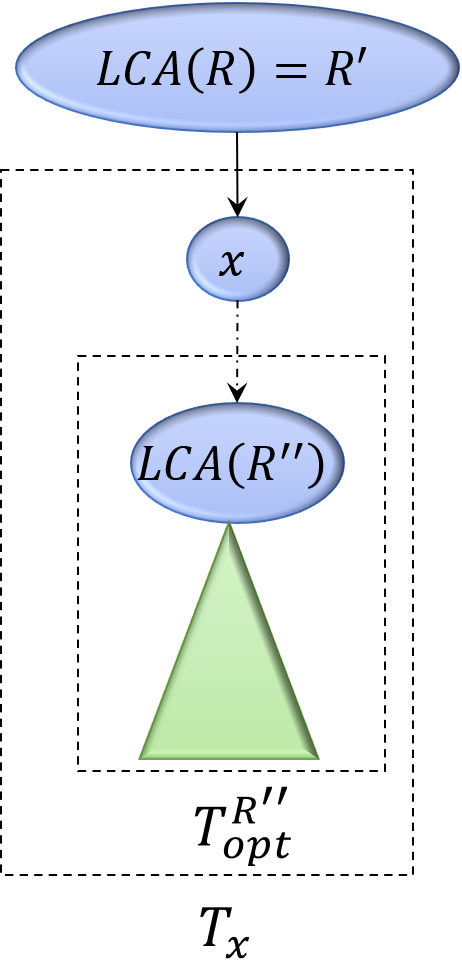}
			\caption{\label{fig:dppa-substruct-c1} When $\lca{\terminalset}$ has one subtree.}
		\end{subfigure}
		\caption{\label{fig:optsubstruct}The structure of the output \sa. $\terminalset'$ and $\terminalset''$ represent any non-empty subset of the input terminal set $\terminalset$. Triangles represent subtrees while circles represent nodes in the arborescence. The dotted and solid arrows represent paths and edges respectively.}
	\end{figure}
	
	Clearly, $\solarb_{x}$ is an optimal \sa ~spanning $\terminalset' \cup \{x\}$; otherwise we can replace $\solarb_{x}$ by another arborescence of smaller cost which (when combined with the rest of $\optarb^{\terminalset}$) would result in a valid arborescence with smaller cost, contradicting the optimality of $\optarb^{\terminalset}$.	
	Now, $\solarb_{x}$ can be split into a path from $x$ to $\lca{\terminalset'}$ and $\optarb^{\terminalset'}$ (it could be that $x = \lca{\terminalset'}$ in which case the path is trivial/empty). 
	Therefore, we have decomposed the edge $(\lca{\terminalset},x)$ and $\solarb_{x}$ into a path from $\lca{\terminalset}$ to $\lca{\terminalset'}$ and $\optarb^{\terminalset'}$.	
	Let $\solarb_{rem}$ be the subtree of $\optarb^{\terminalset}$ obtained by removing the edge $(\lca{\terminalset},x)$ and $\solarb_{x}$ (see Figure~\ref{fig:dppa-substruct-c2}). 
	A similar proof as above can be used to show that $\solarb_{rem}$ is an optimal \sa ~spanning $\terminalset''\cup\{\lca{\terminalset}\}$ and can be decomposed into a path from $\lca{\terminalset}$ to $\lca{\terminalset''}$ and $\optarb^{\terminalset''}$. 
	
	We now consider the case when $\lca{\terminalset}$ has only one child $x$ (see Figure~\ref{fig:dppa-substruct-c1}). 
	$\lca{\terminalset}$ must be a terminal, otherwise $x = \lca{\terminalset}$.
	Let $\terminalset' = \{\lca{\terminalset}\}$ and $\terminalset'' = \terminalset\setminus \terminalset'$. 
	Then $\optarb^{\terminalset}$ can be split into the four subtrees/paths that the lemma requires.
\end{proof}

\autoref{lemma:optsubstruct} leads to the following corollary:
\begin{corollary}
	\label{cor:r-fpt-eqn}
	The cost of an optimal \sa~on the terminal set $\terminalset$ is:
	\begin{equation*}
		\label{eqn:optsubstruct}
		\opt_{\terminalset} = \min\limits_{\phi\neq \terminalset'\subset \terminalset} \Big\{ \opt_{\terminalset'} + \opt_{\terminalset \setminus \terminalset'}	+ \hdist{\lca{\terminalset}, \lca{\terminalset'}} + \hdist{\lca{\terminalset}, \lca{\terminalset \setminus \terminalset'}} \Big\}
	\end{equation*}
\end{corollary}
\autoref{algo:r:fpt} uses \autoref{cor:r-fpt-eqn} to compute the optimal cost on every possible subset of the terminal set, and stores the DP backpointers to the subsets that result in the optimal cost.
Finally, the \sa~is constructed by following these backpointers. 
The pseudocode is given in \autoref{algo:r:fpt} and the algorithm is analyzed in \autoref{thm:r:fpt}.
We comment on the usage of this algorithm for the undirected hypercube case in \autoref{remark:r:fpt}.

\begin{algorithm}[h]
	\caption{\label{algo:r:fpt} A Dynamic Programming Algorithm for the \msap.}
	\AlgInput{Dimension of the hypercube $\dimension$, terminal set $\terminalset$.}\\
	\AlgParameter{The number of input terminals $\modr$.}\\
	\AlgOutput{An optimal Steiner arborescence $\optarb$.}\\
	\AlgDS{A vector $M$ such that $M[S]$ stores the optimal cost of the \sa ~spanning $S\subseteq \terminalset$. There are two backpointers from $S$ to $S'\subset S$ and $S'' = S \setminus S'$. Initially, the values in the vector is set to zero and the backpointers to NULL.}
	\begin{algorithmic}[1]
		\For{$\terminalset'\subseteq \terminalset$ (in increasing order of $|\terminalset'|$)}
			\If{$|\terminalset'| > 1$}
				\State $M[\terminalset']$ is computed using \autoref{cor:r-fpt-eqn} as: 
				\vspace*{-\baselineskip} 
				\begin{equation*}
					M[\terminalset'] \gets \min\limits_{\phi\neq S\subset \terminalset'} \{ M[S] + M[\terminalset' \setminus S] + \hdist{\lca{\terminalset'},\lca{S}} + \hdist{\lca{\terminalset'}, \lca{\terminalset' \setminus S}} \}
				\end{equation*}
				\vspace*{-1.5\baselineskip} 
				\State Let $S_{min} \subset \terminalset'$ be the subset that results in the minimum value. Add the following backpointers -- $\terminalset'$ to $S_{min}$ and $\terminalset'$ to $\terminalset' \setminus S_{min}$.
			\EndIf
		\EndFor
		\State Add $\zerom$ to the arborescence $\optarb$.
		\State Traverse the backpointers from $R$ until NULL (e.g., using a DFS) and add $\mathtt{LCA}$ of the subsets to $\optarb$ as roots of the corresponding subtrees.
	\end{algorithmic}
\end{algorithm}

\rfpt*
\begin{proof}
	The optimality of the output \sa~follows directly from \autoref{lemma:optsubstruct} and \autoref{cor:r-fpt-eqn}. 
	Since the least common ancestor can be computed in $\Oh{\modr\dimension}$ time and there are at most $\modr \dimension$ edges in the optimal \sa, the backtracking can be performed in $\Oh{(\modr \dimension)^2}$ time. 
	Thus, the time complexity of the algorithm is dominated by the time required by the iterations. The loop runs for every possible subset of the terminal set $\terminalset$ and for every subset it explores all its proper subsets. We assume the two table lookups take unit time each. Thus the total time required by the loop is:
	
	\begin{equation*}
		\begin{split}
			&\sum_{|\terminalset'|=2}^{\modr} \binom{\modr}{|\terminalset'|} \sum_{|S|=1}^{|\terminalset'|-1} \binom{|\terminalset'|}{|S|} (2 + |\terminalset'|\dimension + |S|\dimension + (|\terminalset'|-|S|)\dimension + 2\dimension)\\
			&\leq \sum_{|\terminalset'|=0}^{\modr} \binom{\modr}{|\terminalset'|} \sum_{|S|=0}^{|\terminalset'|} \binom{|\terminalset'|}{|S|} 2(1 + |\terminalset'|\dimension + \dimension)		
			\leq 2(1 + |\terminalset|\dimension + \dimension) \sum_{|\terminalset'|=0}^{\modr} \binom{\modr}{|\terminalset'|}  2^{|\terminalset'|}\\
            &= 2(1 + |\terminalset|\dimension + \dimension) 3^{\modr}
		\end{split}
	\end{equation*}
	
	The optimal cost for every possible subset of the terminal set is stored in a $\Oh{2^{\modr}}$ length vector. 
	The maximum cost possible is $\modr \dimension$ which requires $\log {\modr\dimension}$ bits. 
	We assume the two pointers use constant space, thus proving the space complexity.
\end{proof}

\begin{remark}\label{remark:r:fpt}
	It may be noted here that our algorithm can be used for the undirected hypercube case but the resulting arborescence, even though would be a correct solution, may not be optimal.
\end{remark}
\section{\fpt~Algorithms Parameterized by Penalty}
In this section, we present two parameterized algorithms on the penalty parameter. 
We first present a randomized algorithm for the parameterized \msap ~that has some similarities to the `buildNPP' algorithm by Blelloch et al.~\cite{WP-FPT-conf} and Sridhar et al.~\cite{WP-FPT-journal} for undirected hypercubes. 
Following this, we present a novel parameterized approximation algorithm for the \msap ~using a \fpt~algorithm for minimum vertex cover problem.

\subsection{Preliminaries for Algorithms on Penalty $q$}
We first study certain properties of the input characters and structure of the output arborescence that have been used to design our algorithms.
Based on these properties, we also present an additional preprocessing that plays a key role in the exact \fpt ~algorithm (and can be optionally used by the other algorithms).

\subsubsection{Background on Classifications of Characters}
Given an arborescence $\solarb$, a character $u \in \mset$ is a `\emph{good character}' of $\solarb$ if $u$ changes along exactly one edge in $\solarb$; otherwise $u$ is a `\emph{bad character}' of $\solarb$~\cite{WP-FPT-conf,WP-FPT-journal}. 
Let $\good{\solarb}$ and $\bad{\solarb}$ denote the set of all good and bad characters of $\solarb$ respectively. 
Clearly, the penalty of $\solarb$, $\q \geq |\bad{\solarb}|$, as also shown in Blelloch et al.~\cite{WP-FPT-conf} and Sridhar et al.~\cite{WP-FPT-journal}.

Given a set of terminals $\terminalset$, a pair of characters $u,v \in \mset$ is said to conflict with each other if $\exists~t_1,t_2,t_3,t_4 \in \terminalset$ such that $(t_1[u]=0 ~\wedge~ t_1[v] = 0)$ and $(t_2[u]=0 \wedge t_2[v] = 1)$ and $(t_3[u]=1 \wedge t_3[v] = 0)$ and $(t_4[u]=1 \wedge t_4[v] = 1)$~\cite{phylogenyBook}. 
This condition is known as the four-gamete condition.
We say a character $u \in \mset$ is \emph{conflicting} if there is another character $v \in \mset$ such that $u$ and $v$ conflict with each other in $\terminalset$.
Similarly, a character $u \in \mset$ is \emph{isolated} if it does not conflict with any other character. 

Blelloch et al.~\cite{WP-FPT-conf} and Sridhar et al.~\cite{WP-FPT-journal} define the conflict graph as an undirected graph $\cg{\terminalset} = (V,E)$, where $V = \mset$ and $E = \{\{u,v\} \mid \text{$u$ and $v$ conflict with each other} \}$.
As it takes $\Oh{\modr}$ time to determine if two bits conflict with each other, the conflict graph can be constructed in $\Oh{\modr \dimension^2}$ time. 

When the conflict graph does not have any edges, i.e., when all the characters are isolated, the instance is known to satisfy the \emph{perfect phylogeny condition} ~\cite{phylogenyBook}. 
When this condition is satisfied, it is well-known that the \msap~can be solved in polynomial time~\cite{perfectPhylogeny-conf,perfectPhylogeny-journal,perfectPhylogenyCSP,phylogenyBasics}, and all the characters are good in the resulting optimal arborescence.
Thus, the following corollary holds true.
\begin{corollary}
	\label{cor:pp-P-time-solvable}
	If there are no edges in the conflict graph, then $\qopt=0$ and the \msap~can be solved in polynomial time.
\end{corollary}

\vspace*{\baselineskip}
\emph{For a character $u \in \mset$, we define $\terminalset_{u} = \{t \in \terminalset \mid t[u] = 1\}$. In an arborescence $\solarb$, we use the notation $e_{u} = (x_{\bar{u}}, x_u)$ to denote an edge along which $u$ changes in $\solarb$ (i.e., $x_{\bar{u}}$ and $x_u$ have the same bit value for all characters except the character $u$ at which $x_{\bar{u}}[u] = 0$ and $x_u[u] = 1$). Note that if $u \in \good{\solarb}$, then there is only one such edge $e_u$.}

\subsubsection{Structure of the Output Arborescence}
We further study the relationship between the properties of characters and structure of output arborescences, as it forms the basis of our algorithms. 
We first show the following bounds on $\qopt$.
\begin{proposition}
	\label{claim:good-char-in-opt-sa}
	For an instance $I$ of the \msap, the following statements hold true.
	\begin{enumerate}
		\item If $\qopt < \dimension$, then every $\optarb$ has at least one good character.
		\item If $\qopt \geq \dimension$, then $\modr \leq 2\qopt + 1$.
	\end{enumerate}
\end{proposition}
\begin{proof}
	If $\qopt < \dimension$, $\opt = \dimension + \qopt < 2\dimension$.
	Then all the characters cannot flip at least twice, which implies that there is at least one character which flips once. In fact, $\optarb$ has at least $m-\qopt$ good characters, since   $m=|\good{\optarb}|+|\bad{\optarb}|$, and $|\bad{\optarb}| \leq \qopt$.   
	If $\qopt \geq \dimension$, $\opt \leq 2\qopt$.
	As $\modr - 1 \leq \opt$, $\modr \leq 2\qopt + 1$.
\end{proof}
So, if $\qopt \geq \dimension$, then the input size is bounded by $\qopt$ and any (trivial) exponential algorithm will construct a \sa ~in \fpt ~time.
Instances for which $\qopt < \dimension$ are more interesting since the hardness of the parameterized version of the problem for this scenario cannot be inferred directly.
Henceforth, we assume $\qopt < \dimension$ and study this case in more details.

We use \autoref{claim:flipping-characters} on the structure of an optimal arborescence multiple times in the design of the exact algorithms.
\begin{proposition}
	\label{claim:flipping-characters}
	Consider a path $P = (x_1, x_2, \ldots, x_k, x_{k+1})$ (for $k \ge 2$) such that $\forall~ i \in [2,k]$ $x_i$ is a Steiner node and has exactly one child (i.e., $x_{i+1}$) in an optimal \sa ~$\optarb$.
	Let the characters $u_1, \ldots, u_k$ flip along the corresponding edges in $P$ in the same order, and $u_1', \ldots, u_k'$ be any permutation of these characters.
	Then there exists an optimal \sa ~$\optarb'$ such that $u_1', \ldots, u_k'$ flip along corresponding edges in the path $P' = (x_1,x_2', \ldots, x_k', x_{k+1})$ in $\optarb'$ in the same order and the rest of $\optarb'$ is identical to $\optarb$.
\end{proposition}
\begin{proof} 
	Observe that $\forall~ j \in [1,k]~ x_1[u_j] = 0$ and $x_{k+1}[u_j] = 1$.
	So there is a path $P' = (x_1,x_2', \ldots, x_k', x_{k+1})$ in $\Qm$ such that $u_1', \ldots, u_k'$ change along the corresponding edges in the same order.
	Also notice that $\forall~ i \in [2,k]$ $x_i' \notin \vertexset{\optarb}$; otherwise if for some $i \in [2,k]$ $x_i'$ is in $\optarb$, then we can replace the path $P$ with a path from $x_i'$ to $x_{k+1}$ and obtain a valid \sa ~with cost reduced at least by $1$.
	As a consequence of these two observations, we can replace the path $P = (x_1, x_2, \ldots, x_k, x_{k+1})$ in $\optarb$ with $P' = (x_1, x_2', \ldots, x_k', x_{k+1})$ and obtain a valid \sa ~$\optarb'$.
	As the length of the new replaced path is same as the length of the original path, $\cost{\optarb'} = \cost{\optarb}$ and $\optarb'$ is also optimal.
\end{proof}

\begin{remark}
	\label{remark:cor-3-proof}
	The above proposition leads to an interesting observation.
	Let $u$ be a good character of $\optarb$, and $\solarb_{u}$ be the subtree of $\optarb$ rooted at $x_{u}$.	
	Now, every terminal $t \in \terminalset_{u}$ is in $\solarb_{u}$.
	Also $\terminalset_{u}$ is rooted at $y = \lca{\terminalset_{u}}$ which must be a descendent of $x_u$.
	Notice that, there cannot be any subtrees along the path from $x_{u}$ to $y$ in $\optarb$.
	Let $u_1, u_2, \ldots, u_{k-1}, u_k$ flip along the path from $x_{\bar{u}}$ to $y$ (here $u_1 = u$) in the same order.
	By \autoref{claim:flipping-characters}, we permute the characters as $u_k, u_2, \ldots, u_{k-1}, u_1$, (i.e., the positions of characters $u_1$ and $u_k$ are swapped), and the resulting arborescence is also optimal (and has the same set of good characters as $\optarb$). 
\end{remark}
The above \autoref{remark:cor-3-proof} leads to the following corollary. 
\begin{corollary}
    \label{cor:x_u=lca(R_u)}
    Consider the \msap ~instance $I = \instance{\dimension, \terminalset}$.
    Let $u$ be a good character of $\optarb$. 
    Then, there exists an optimal \sa ~$\optarb'$ such that $\good{\optarb'} = \good{\optarb}$ and $x'_{u} = \lca{\terminalset_{u}}$. 
    Here $x'_{u}$ denotes the head of the edge ($e'_{u} = (x'_{\bar{u}}, x'_{u})$) along which $u$ changes in $\optarb'$.
\end{corollary}
 
\subsubsection{Additional Preprocessing Step}
We use \autoref{claim:isolated-optimality} to show that isolated characters can be preprocessed using \autoref{lemma:isolated-char-preprocess}.
\begin{proposition}
	\label{claim:isolated-optimality}
	Consider an isolated character $u$. For any optimal \sa ~$\optarb$, $u \in \good{\optarb}$.
\end{proposition}
\begin{proof}
	If $u$ is not a good character in $\optarb$, then $u$ changes along at least two distinct edges $e_1 = (x_1,y_1)$ and $e_2 = (x_2, y_2)$ in $\optarb$. We will now inspect the structure/edges in $\optarb$ to prove this result by contradiction. 
	Let $x$ be that node in $\optarb$ such that the subtree rooted at $x$ contains both $e_1$ and $e_2$ (but no subtree rooted at any descendant of $x$ contains both $e_1$ and $e_2$).
	Note that there is a terminal $t_1$ and $t_2$ in subtrees $T_1$ and $T_2$ rooted at $y_1$ and $y_2$ respectively due to the optimality condition.
	
	A subtree is said to be along a path if it is rooted at a node in the path with outdegree at least two and contains at least one terminal; it may also be the case that a node is itself a terminal and hence is consider as a subtree rooted at the node consisting of that single terminal only.
	Let $P_1$ and $P_2$ be the paths from $x$ to $x_1$ and from $x$ to $x_2$ respectively.
	We first address the case when there are no subtrees along $P_1$ and $P_2$. 
	Then by \autoref{claim:flipping-characters}, the order in which the characters change in both $P_1$ and $P_2$ is irrelevant.
	So, we can add a new node $x_u$, change $u$ first along the edge $(x,x_u)$ and all the other characters can change along edges in the subtree rooted at $x_u$.
	This results in a valid \sa~with cost reduced at least by $1$, contradicting the optimality criteria.
	Hence there must be at least one terminal $t_3$ in some subtree ($T_3$) along $P_1$ or $P_2$.
	
	Let $B_1$ and $B_2$ be the set of characters that change in $P_1$ and $P_2$ respectively.
	If $B_1 = B_2$, then $x_1 = x_2$, which contradicts the claim that $e_1$ and $e_2$ are distinct.
	Let $T_3$ be rooted at some node $z$ in $P_1$.
	If there is a character $v \in B_1 \setminus B_2$ that flips along some edge in the path from $x$ to $z$, then $(t_1[u] = 1 \wedge t_1[v] = 1)$ and $(t_2[u] = 1 \wedge t_2[v] = 0)$ and $(t_3[u] = 0 \wedge t_3[v] = 1)$.
	Now the root $r = \zerom$ has $(r[u] = 0 \wedge r[v] = 0)$, implying that $u$ conflicts with $v$, which is a contradiction.
	
	We now consider the case when there are no such characters along both the paths.
	Let $v_1 \in B_1$ and $v_2 \in B_2$ be the first characters that flip along some edge in $P_1$ and $P_2$ respectively such that $v_1 \notin B_2$ and $v_2\notin B_1$. 
	Let $v_1$ and $v_2$ change along the edges $(x_{\bar{v_1}}, x_{v_{1}})$ and $(x_{\bar{v_2}}, x_{v_{2}})$.
	Observe that there are no subtrees along the paths $P_1'$ (from $x_{v_{1}}$ to $x_1$) and $P_2'$ (from $x_{v_{2}}$ to $x_2$).
	Again by \autoref{claim:flipping-characters}, the characters changing along the paths from $x_{\bar{v_{1}}}$ to $x_1$ and from $x_{\bar{v_{2}}}$ to $x_2$ can be permuted to obtain an optimal \sa.
	Let $B_1', B_2' \subseteq B_1 \cap B_2$ be the subset of characters that change along $P_1'$ and $P_2'$ respectively.
	Then, from $x_{\bar{v_1}}$ we first change all characters in $B_1'$ and then we change $v_1$. Similarly, from $x_{\bar{v_2}}$ we first change all characters in $B_2'$ and then we change $v_2$. 
	In the new paths, let $v_1$ and $v_2$ change along the edges $(x_{\bar{v_1}}', x_{v_{1}}')$ and $(x_{\bar{v_2}}', x_{v_{2}}')$ respectively. Then, $x_{\bar{v_{1}}}' = x_{\bar{v_{2}}}'$, contradicting the optimality of $\optarb$.
\end{proof}

We use \autoref{claim:isolated-optimality} to prove the following lemma.
\begin{lemma}
	\label{lemma:isolated-char-preprocess}
	Let $u \in [\dimension]$ be an isolated character in a \msap~instance $I = \instance{\dimension, \terminalset}$. Let $I' = \instance{\dimension', \terminalset'}$ be defined as $[\dimension'] = [\dimension] \setminus \{u\}$ and $\terminalset' = \{t' \mid \exists~t \in \terminalset \text{ such that } \forall~ v\in [\dimension']~ t'[v] = t[v]\}$. 
	Then, $\opt_{\terminalset} = \opt_{\terminalset'} + 1$.
\end{lemma}
\begin{proof}
	By \autoref{claim:isolated-optimality}, there is exactly one edge $e_u = (x_{\bar{u}}, x_u)$ along which $u$ changes in $\optarb^{\terminalset}$.
	We remove the edge $e_u$, connect every child of $x_u$ in $\optarb^{\terminalset}$ to $x_{\bar{u}}$ and delete the character $u$ from every node.
	The resulting arborescence is a valid solution of $I'$, and hence $\opt_{\terminalset'} \le \opt_{\terminalset} - 1$.	
	
	To prove the other bound $\opt_{\terminalset'} \ge \opt_{\terminalset} - 1$, we add the character $u$ to every node of $\optarb^{\terminalset'}$ and set it to $0$.	 
	Let the resulting arborescence be $\solarb$.
	For every terminal $t \in \terminalset_{u}$, let $\bar{t}$ be defined as:
	\[\bar{t}[v] = 
	\begin{cases}
		0, & if~v=u\\
		t[v], & otherwise
	\end{cases}
	\]
	Let $\bar{\terminalset_{u}} = \{\bar{t} \mid t \in \terminalset_{u}\}$.
	Note that $\solarb$ is an optimal \sa ~for the terminal set $\terminalset \cup \bar{\terminalset_{u}} \setminus \terminalset_{u}$, and the subset $\bar{\terminalset_{u}}$ is rooted at $y_{\bar{u}} = \lca{\bar{\terminalset_{u}}}$ in $\solarb$.
	
	Let the character $w$ change along an edge $e = (y_{\bar{u}}, x)$ in $\solarb$; and let the subtree $\solarb_{x}$ rooted at $x$ contains a terminal from $\bar{\terminalset_{u}}$. 
	Then $\solarb_{x}$ does not contain any terminal $t' \notin \bar{\terminalset_{u}}$, otherwise $u$ will conflict with $w$.
	We do the following modifications to $\solarb$ to obtain $\solarb'$.
	\begin{itemize}
		\item Add the edge $(y_{\bar{u}}, y_{u})$ that changes on $u$.
		\item For every edge $e = (y_{\bar{u}}, x)$ in $\solarb$ such that $\solarb_x$ contains terminals from $\bar{\terminalset_{u}}$, remove the edge $e$ and add the edge $(y_u,x)$.
		\item For every node $x$ in the subtree rooted at $y_u$, flip $x[u]$ to $1$.
	\end{itemize}
	Trivially, $\solarb'$ is a valid \sa~on $\terminalset$.
\end{proof}

Observe that every character of an instance that has been preprocessed by \autoref{lemma:isolated-char-preprocess} is a conflicting character.
Hence, the following corollary holds true by \autoref{claim:good-char-in-opt-sa}.
\begin{corollary}
	\label{cor:cc<=q_=>_|R|<=2q+1}
	In every instance preprocessed by \autoref{lemma:normalization} and \autoref{lemma:isolated-char-preprocess}, if number of conflicting characters $\leq \qopt$, then $\modr \leq 2\qopt + 1$.
\end{corollary}

\subsection{Randomized \fpt~Algorithm for the \msap}
For this algorithm, we exploit another property of the good characters in \autoref{claim:q:fpt:good-char}, and use \autoref{cor:x_u=lca(R_u)} to prove the same.

\begin{proposition}
	\label{claim:q:fpt:good-char}
	Let $\optarb^{(0)}$ be an optimal \sa ~of the \msap ~instance $I = \instance{\dimension, \terminalset}$, and let $u_1, \ldots, u_{\kappa} \in \good{\optarb^{(0)}}$.
	Then, there exists an optimal \sa ~$\optarb^{(\kappa)}$ of $I$ such that $\good{\optarb^{(\kappa)}} = \good{\optarb^{(0)}}$, and for every $i \in [1, \kappa]$, $x_{u_{i}}^{(\kappa)} = \lca{\terminalset_{u_{i}}^{(\kappa,i)}}$.
    Here $x_{u_{i}}^{(\kappa)}$ denotes the head of the edge ($e_{u_{i}}^{(\kappa)} = (x_{\bar{u}_{i}}^{(\kappa)}, x_{u_{i}}^{(\kappa)})$) along which the good character $u_i$ changes in $\optarb^{(\kappa)}$; and $\terminalset_{u_i}^{(\kappa,i)} = \{t \in \terminalset^{(\kappa,i)} \mid t[u_i] = 1 \}$, with $\terminalset^{(\kappa,i)} = \terminalset ~\cup~ \bigcup_{j \in [1,i-1]} \{x_{\bar{u}_j}^{(\kappa)}, x_{u_{j}}^{(\kappa)}\}$.
\end{proposition}
\begin{proof}
	We prove the statement using induction on $\kappa$.
	For the base case, $\terminalset^{(1,1)} = \terminalset \implies \terminalset_{u_{1}}^{(1,1)} = \terminalset_{u_{1}}$, and the statement holds true by \autoref{cor:x_u=lca(R_u)}.
	
	Let $\optarb^{(i-1)}$ be the optimal \sa ~spanning $\terminalset$ with the same set of good characters as $\optarb^{(0)}$ such that the above statement holds true for the characters $u_1, \ldots, u_{i-1}$. In $\optarb^{(i-1)}$, let the character $u_i$ change along the edge $e_{u_{i}}^{(i-1)} = (x_{\bar{u}_{i}}^{(i-1)}, x_{u_{i}}^{(i-1)})$; also define $\terminalset^{(i-1,i)}$ and $\terminalset_{u_i}^{(i-1,i)}$ as in the proposition statement.
    Notice that $\terminalset^{(i-1,i)}$ is $\terminalset$ augmented with possibly other nodes in $\optarb^{(i-1)}$, so $\optarb^{(i-1)}$ is also an optimal solution for spanning $\terminalset^{(i-1,i)}$.
	
	We construct $\optarb^{(i)}$ from $\optarb^{(i-1)}$ as follows:
	\begin{enumerate}
		\item[Case 1:] If $x_{u_{i}}^{(i-1)} = \lca{\terminalset_{u_{i}}^{(i-1,i)}}$, then $\optarb^{(i)} \gets \optarb^{(i-1)}$.
		\item[Case 2:] Otherwise, $\optarb^{(i)} \gets $ modified $\optarb^{(i-1)}$ as per \autoref{remark:cor-3-proof}.
	\end{enumerate}
	
	Clearly in both cases, $\optarb^{(i)}$ satisfies the proposed statement for $u_i$ (i.e., $x_{u_{i}}^{(i)} = \lca{\terminalset_{u_{i}}^{(i,i)}}$), while retaining the same set of good characters, arborescence cost and hence optimality of spanning $\terminalset$ as $\optarb^{(i-1)}$ (by \autoref{remark:cor-3-proof} and proof of \autoref{claim:flipping-characters}). 
    To complete the proof that $\optarb^{(i)}$ satisfies the proposed statement for every character $u_1, \ldots, u_i$, we prove the claim: $\forall j \in [1, i-1]$, the edge along which $u_j$ changes is the same between $\optarb^{(i)}$ and $\optarb^{(i-1)}$ (thereby implying $\terminalset^{(i,j)} = \terminalset^{(i-1,j)}$ and hence $x_{u_{j}}^{(i)} = \lca{\terminalset_{u_{j}}^{(i,j)}}$). 
    
	The above claim holds trivially in Case 1. In Case 2, consider the path $P$ in $\optarb^{(i-1)}$ from $x_{\bar{u}_{i}}^{(i-1)}$ to $\lca{\terminalset_{u_{i}}^{(i-1,i)}}$, with $v$ being the last character that changes along this path.
	For every $j \in [1, i-1]$, $u_j$ does not change along any edge in $P$ (otherwise, it contradicts the last node in $P$ being $\lca{\terminalset_{u_{i}}^{(i-1,i)}}$); so, although the positions of $u_i$ and $v$ along $P$ are swapped to construct $\optarb^{(i)}$ in \autoref{remark:cor-3-proof}, the edge along which $u_j$ changes stays invariant, thus proving the above claim. 
\end{proof}

\subsubsection{Algorithm Outline}
Our proposed algorithm iteratively constructs an optimal \sa ~$\optarb$ with one good character (and its associated edge) at a time. 
The algorithm maintains a partition $\mathcal{P}$ of the set of input terminals and other augmented terminals. 
During an iteration, each subset in $\mathcal{P}$ is analyzed to obtain its conflicting characters; and from the union of conflicting characters across all the subsets, the algorithm chooses one character $u$ uniformly at random. 
The cardinality of this union is referred to as the {\it ``total number of conflicting characters''} in this work.

Assume that the random choice $u$ during any iteration leads to a successful outcome, viz., $u \in \good{\optarb}$. 
Then, the edge along which $u$ changes in $\optarb$ is constructed using \autoref{claim:q:fpt:good-char}. 
This edge should split $\optarb$ into two subtrees, each covering a non-empty subset of the terminals. 
The terminal set is partitioned into these two subsets (along with the endpoints of the newly added edge) and the entire process is repeated until the total number of conflicting characters is at most $q$. 
Exact details/pseudocode is in \autoref{algo:q:fpt}. 

Observe that any subset newly created in an iteration is not preprocessed and hence may contain invariant or isolated characters. 
So, after the loop terminates, every remaining subset is preprocessed by \autoref{lemma:normalization} and \autoref{lemma:isolated-char-preprocess} to remove such characters and the optimal \sa ~of the subset is constructed independently using \autoref{algo:r:fpt}.
It may be noted here that since by \autoref{claim:good-char-in-opt-sa} and \autoref{cor:cc<=q_=>_|R|<=2q+1}, the preprocessed subset with at most $q$ (conflicting) characters has at most $2q + 1$ terminals, the optimal \sa ~can be constructed in \fpt ~time.
Analysis of the overall algorithm is in \autoref{thm:q:fpt}.

\begin{algorithm}[h]
	\caption{\label{algo:q:fpt} Randomized Algorithm for the Parameterized \msap.}
	\AlgInput{Dimension of the hypercube $\dimension$, terminal set $\terminalset$, an integer $q\geq0$.}\\
	\AlgParameter{Penalty on the cost $q$.}\\
	\AlgOutput{An optimal Steiner arborescence $\optarb$.}\\
    \AlgDS{A partition $\mathcal{P}$ of the set of input and newly augmented terminals.}
	\begin{algorithmic}[1]
		\State Initialize $\mathcal{P}$ with the trivial partition $\{\terminalset\}$.
		\While{total number of conflicting characters $> q$}
			\State From the union of conflicting characters across all subsets in $\mathcal{P}$, choose a character $u$ uniformly at random. Let $S$ be a subset in which $u$ is conflicting.
			\State Partition $S$ into $S_{u} = \{t \in S \mid t[u] = 1\}$ and $S_{\bar{u}} = S \setminus \{S_{u}\}$. Let $x_{u} \gets \lca{S_{u}}$. 
            \State Add edge $e_{u} = (x_{\bar{u}}, x_{u})$, which changes on $u$, to $\solarb$.
			\State Update $\mathcal{P}$ as $\mathcal{P} \gets \mathcal{P} \cup \{S_{u} \cup \{x_{u}\}, S_{\bar{u}} \cup \{x_{\bar{u}}\}\} \setminus \{S\}$.
   			\If{number of iterations $> q$}
				\State \Return error
			\EndIf
		\EndWhile
		\ForAll{remaining subsets $S$ in $\mathcal{P}$}
			\State Preprocess $S$ by \autoref{lemma:normalization} and \autoref{lemma:isolated-char-preprocess}.
			\State Use \autoref{algo:r:fpt} to construct its optimal \sa~$\optarb^S$.
			\State $\solarb \gets \solarb \cup \optarb^S$.
		\EndFor
	\end{algorithmic}
\end{algorithm}

\qfpt*
\begin{proof}
	Consider an optimal \sa ~$\optarb$.
	By \autoref{claim:good-char-in-opt-sa}, there must be a conflicting character that is good in $\optarb$ if the loop is invoked. Assume that the algorithm's randomly chosen characters $u_1, \ldots, u_{\kappa}$ are good characters of $\optarb$ (success probability has been calculated below).
    Then, by \autoref{claim:q:fpt:good-char}, there is an optimal \sa~ $\optarb^{(\kappa)}$ with the same set of good characters as $\optarb$ and in which $\forall i \in [1, \kappa]$, $x_{u_{i}}^{(\kappa)} = \lca{\terminalset_{u_{i}}^{(\kappa,i)}}$. Recall that the good character $u_i$ in $\optarb^{(\kappa)}$ changes along the edge $e_{u_{i}}^{(\kappa)} = (x_{\bar{u}_{i}}^{(\kappa)}, x_{u_{i}}^{(\kappa)})$.

    \underline{\emph{Optimality:}}
    Under the assumption stated above that all chosen characters $u_1, \ldots, u_{i}$ are good characters of $\optarb$ or equivalently $\optarb^{(\kappa)}$, the character $u_i$, chosen during iteration $i$, will be conflicting in exactly one subset $S$ and invariant in all other subsets.
    Then $\opt_{S} = \opt_{S_{\bar{u}_i} \cup \{x_{\bar{u}_i}^{(\kappa)}\}} + \opt_{S_{u_{i}} \cup \{x_{u_{i}}^{(\kappa)}\}} + 1$. 
    This validates dividing the subproblem $S$ into two other subproblems using $x_{u_{i}}$.

    To find $x_{u_{i}}$, note that $\lca{S_{u_{i}}} = \lca{\terminalset_{u_{i}}^{(\kappa,i)}}$    
    (because $\terminalset_{u_{i}}^{(\kappa,i)}$ is a set of input/augmented terminals in the subtree $T'$ of $\optarb^{(\kappa)}$ rooted at $x_{u_{i}}^{(\kappa)}$, and $S_{u_i} \subseteq \terminalset_{u_{i}}^{(\kappa,i)}$ is the set of input/augmented terminals in $T'$ after pruning subtrees in $T'$, none of which affect the LCA computation; in detail, $\forall j \in [1,i-1]$ such that $x_{u_j}$ is a descendant of $x_{u_i}$ in $\optarb^{(\kappa)}$, the pruning removes the subtree of $T'$ rooted at $x_{u_{j}}$ successively). 

	Thus, the algorithm generates $\optarb^{(\kappa)}$, if the chosen vertices are all good characters of $\optarb$, $x_{u_{i}}^{(\kappa)}$ is set to $\lca{S_{u_{i}}}$ and an exact algorithm is used to construct an optimal \sa~on each of the remaining subsets. 

    \underline{\emph{Loop Convergence:~\cite{WP-FPT-conf,WP-FPT-journal}}}	
	Let the subtrees spanning $S$, $S_{u_{i}} \cup \{x_{u_{i}}^{(\kappa)}\}$ and $S_{\bar{u}_i} \cup \{x_{\bar{u}_i}^{(\kappa)}\}$ in $\optarb^{(\kappa)}$ be $\solarb$, $\solarb_{u_{i}}$ and $\solarb_{\bar{u}_{i}}$ respectively. 
	Since $u_i$ is a conflicting character, there exists at least one character $v$ that conflicts with $u_i$ in $S$; and thus $v$ changes at least once in both $\solarb_{u_{i}}$ and $\solarb_{\bar{u}_{i}}$.
	If $q(v)$, $q_{u_{i}}(v)$ and $q_{\bar{u}_{i}}(v)$ denotes the penalty incurred due to $v$ (i.e., number of edges along which $v$ changes $-1$) in $\solarb$, $\solarb_{u_{i}}$ and $\solarb_{\bar{u}_{i}}$ respectively, then $(q_{u_{i}}(v)+1) + (q_{\bar{u}_{i}}(v)+1) = (q(v)+1)$. This implies 
	$q_{\solarb_{u_{i}}} + q_{\solarb_{\bar{u}_{i}}} + 1 \leq q_{\solarb}$.
    
	So if all choices are correct, we are reducing the total penalty (sum of penalty of subtrees of $\optarb^{(\kappa)}$ spanning each subset in the current partition) by at least $1$ with each iteration. Given that $\qopt \leq q$, there can be at most $q$ iterations (otherwise, total penalty will become 0, and there will be no  conflicting characters).
	
	\underline{\emph{Success Probability:}}
	We now compute the probability that all the chosen characters are good characters of $\optarb$.
    Let the loop runs $\kappa$ times ($\kappa \leq q$).    
	Then before the start of the first iteration there are at least $q + \kappa$ conflicting characters, out of which at most $q$ are bad characters in $\optarb$.
	With each iteration, we are reducing the number of characters by at least $1$, implying that before the start of $j^{th}$ iteration, there must be at least $(q + \kappa) - (j - 1)$ conflicting characters.
	Thus, 
	\begin{equation*}
		\begin{split}
			&\Pr[ u_j \in \good{\optarb} \mid u_{j-1},\ldots,u_{1} \in \good{\optarb} ] \geq 1 - \frac{q}{q + \kappa - (j-1)} 
			= \frac{\kappa - j + 1}{q + \kappa - j + 1}\\
			&\Pr[\text{all chosen characters are good}] = \Pr[u_{\kappa},\ldots,u_1 \in \good{\optarb}] \\
			&= \Pr[u_{\kappa} \in \good{\optarb} \mid u_{\kappa-1},\ldots,u_{1} \in \good{\optarb} ] \ldots \Pr[u_1 \in \good{\optarb}]\\
			&\geq \prod_{j=1}^{\kappa} \frac{\kappa - (j-1)}{q + \kappa - (j-1)}
			= \frac{\kappa! q!}{(q + \kappa)!} 
			= \frac{1}{\binom{q + \kappa}{\kappa}} 
			\geq \frac{1}{2^{q + \kappa}}
			\geq \frac{1}{4^q}
		\end{split}
	\end{equation*}
	
	\emph{\underline{Time Complexity:}}
	Every run of the iteration takes $\Oh{\modr \dimension^2}$ time.
	In each iteration we are partitioning exactly one subset into two, and we have at most $q$ iterations.
	Hence we can have at most $q + 1$ subsets remaining at the termination of the iteration.
	In any preprocessed subset $S$, there are at most $q$ conflicting characters (termination condition of loop) and hence at most $2q + 1$ terminals (by \autoref{cor:cc<=q_=>_|R|<=2q+1}).
	So, \autoref{algo:r:fpt} takes $\Oh{3^{2q + 1} (2q+1)q + ((2q+1)q)^2} = \Oh{9^q q^2}$ time.
	Thus, construction of the optimal \sa~on all the remaining subsets require $\Oh{9^q q^2 (q+1)}$ time.
\end{proof}

\subsubsection{Derandomization}
To derandomize \autoref{algo:q:fpt}, we propose a method inspired from the derandomization algorithm for the undirected version of the problem \cite{WP-FPT-conf,WP-FPT-journal}. Consider an optimal \sa ~$\optarb$ for an instance $I$, preprocessed by \autoref{lemma:normalization} and \autoref{lemma:isolated-char-preprocess}. 
Every conflicting character in $I$ is either a good character or a bad character in $\optarb$ -- our proposed algorithm uses a bounded search tree that branches on these two choices up to a maximum depth of $2q$. 
This algorithm's pseudocode is in \autoref{algo:q:fpt:new} and analysis is in \autoref{lemma:q:fpt:new}.

\begin{algorithm}[!]
	\caption{\label{algo:q:fpt:new} Deterministic Algorithm for the Parameterized \msap.}
	\AlgInput{Dimension of the hypercube $\dimension$, terminal set $\terminalset$, an integer $q\geq0$.}\\
	\AlgParameter{Penalty on the cost $q$.}\\
	\AlgOutput{An optimal Steiner arborescence $\optarb$.}\\
	\AlgDS{A depth-bounded search tree (DBST) rooted at $N_0$. Every node $N$ in the DBST stores a partition $\mathcal{P}(N)$ of the set of input terminals and other augmented terminals; and two sets for bookkeeping -- $G(N)$ and $B(N)$. Every leaf node $N$ in the DBST is marked either ``valid'' or ``pruned'' at the end of the algorithm.}
	\begin{algorithmic}[1]
		\State $\mathcal{P}(N_0) \gets \{\terminalset\}$, $G(N_0) \gets \phi$ and $B(N_0) \gets \phi$.
		\While{there is an unmarked leaf node $N$ at a depth $\leq 2q$}
            \If{$|G(N)| > q$ or $|B(N)| > q$}
                \State Mark $N$ as ``pruned''.
            \ElsIf{total number of conflicting characters $\leq q$}
                \State Mark $N$ as ``valid''.
            \ElsIf{depth of the node $N$ is $2q$}
                \State Mark $N$ as ``pruned''.
            \Else
                \State Choose a conflicting character $u$ arbitrarily and let $S \in \mathcal{P}(N)$ be a subset in which $u$ is conflicting.  
	   	   \State Branch $N$ into two unmarked nodes $N_1$ and $N_2$.
                \State At $N_1$, $G(N_1) \gets G(N)$ and $B(N_1) \gets B(N) \cup \{u\}$ and $\mathcal{P}(N_1) \gets \mathcal{P}(N)$.
                \State At $N_2$, $G(N_2) \gets G(N) \cup \{u\}$, $B(N_2) \gets B(N)$ and $\mathcal{P}(N_2) \gets \mathcal{P}(N) \cup \{S_u \cup \{x_u\}, S_{\bar{u}} \cup \{x_{\bar{u}}\}\} \setminus \{S\}$ (cf. \autoref{algo:q:fpt} Lines 4--6).
            \EndIf
        \EndWhile
		\For{every leaf node $N$ marked as ``valid"}
            \State Construct a \sa ~as per \autoref{algo:q:fpt} Lines 11--15.
        \EndFor

        \If{there are no ``valid" leaf nodes or every arborescence has cost $> \dimension + q$}
            \State \Return No-Instance.
        \Else
            \State \Return the \sa ~with minimum cost.
        \EndIf
	\end{algorithmic}
\end{algorithm}

\begin{lemma}
    \label{lemma:q:fpt:new}
    The \msap ~can be solved in $\Ohs{36^q}$ time.
\end{lemma}
\begin{proof}
	\emph{\underline{Correctness:}}
    The algorithm cannot return a Steiner arborescence of cost at most $\dimension+q$ for a no-instance. 
    So it suffices to consider a yes-instance $I$, and an optimal arborescence $\optarb$ of $I$. 
    Call a node $N$ in the depth-bounded search tree (DBST) as ``correct'' if the algorithm's choices so far until this node are correct, i.e., if $G(N) \subseteq \good{\optarb}$ and $B(N) \subseteq \bad{\optarb}$. 
    Then, the algorithm is correct if there exists a ``correct'' leaf node $N'$ in the DBST that is also marked ``valid'' -- the arborescence constructed at this node will yield the desired solution with cost at most $\dimension + q$.
    
    We first prove the existence of a correct leaf node. 
    The root node of the DBST is trivially correct. 
    If the algorithm doesn't split $N$ further, then $N$ itself is the correct leaf. 
    Otherwise, let $u$ be the conflicting character chosen by the algorithm to split $N$. 
    Since $u$ is either a good or a bad character in $\optarb$, and we branch out and explore both these options in the DBST, one of the two children ($N_c$) of $N$ is also correct. 
    Repeating this argument (for at most $2q$ iterations), the correct leaf is either $N_c$ itself, or a descendant of $N_c$. 
    This proof also shows that each ancestor node of this correct leaf is also correct.  

    We now show that the correct leaf, denoted $N'$, is marked valid in \autoref{algo:q:fpt:new}, Line 6. 
    Since $B(N') \subseteq \bad{\optarb}$, $|B(N')| \leq q$. 
    Since $G(N') \subseteq \good{\optarb}$, we  can prove that $|G(N')| \leq q$ by contradiction as follows. 
    Assume $|G(N')| > q$, and let $N''$ be the ancestor of $N'$ such that $|G(N'')| = q$.
    Then, the total number of conflicting characters will have become $0$ at $N''$ (cf. ``Loop Convergence'' section in the proof of \autoref{thm:q:fpt}), and hence $N''$ would not have been split further. 
    Finally, we claim the total number of conflicting characters at $N'$ is at most $q$ -- otherwise, $N'$ would've been split further. 
    The depth bound of $2q$ won't prevent this split, as the algorithm chooses $q$ bad characters of $\optarb$ in the worst case and has at least $q$ other iterations to choose a sufficient number of good characters to bring down the total number of conflicting characters to at most $q$.
    
    \emph{\underline{Time Complexity:}}
    As the search tree has a maximum depth of $2q$ and branches into exactly two nodes from every node, there are at most $2^{2q+1} - 1$ nodes in the tree and each node takes $\Oh{\modr \dimension^2}$ time.
    There can be at most $2^{2q}$ leaf nodes in the search tree, each consisting of at most $q+1$ subsets in its partition. Further, in a ``valid'' leaf's partition, every subset $S$ has at most $q$ conflicting characters and therefore at most $2q + 1$ terminals.
    Thus, \autoref{algo:r:fpt} takes $\Ohs{3^{2q+1}(q+1)}$ time to construct the optimal \sa s of the subsets in a valid leaf's partition.
\end{proof}
\subsection{Parameterized Approximation based on Vertex Cover}
\label{subsec:q:approx}
To design an approximation algorithm related to the optimal penalty $\qopt$, we first explore a lower bound on $\qopt$ using the minimum vertex cover of the conflict graph.

\subsubsection{A Relationship between Penalty and Minimum Vertex Cover}
Observe that if characters $u$ and $v$ conflict with each other in $\terminalset$, then in every \sa ~$\solarb$, $u$ and $v$ both cannot be good characters of $\solarb$.
We use this to we relate the number of bad characters in any arborescence $\solarb$ to the vertex cover of the conflict graph in \autoref{claim:vc-cg-relation}.
\begin{proposition}
	\label{claim:vc-cg-relation}
	For every \sa~$\solarb$ of $\terminalset$, $\bad{\solarb}$ is a vertex cover of $\cg{\terminalset}$. Further, for every minimal vertex cover $\vc$ of $\cg{\terminalset}$,  there is a \sa ~$\solarb$ such that $\bad{\solarb} = \vc$.
\end{proposition}
\begin{proof}
	Let $\{u,v\}\in \edgeset{\cg{\terminalset}}$.
	Then there are terminals $t_1, t_2, t_3$ such that $t_1[u] = 0 \wedge t_1[v] = 1$, $t_2[u] = 1 \wedge t_2[v] = 0$ and $t_3[u] = 1 \wedge t_3[v] = 1$.
	Let $u$ change exactly once in $\solarb$, along the edge $e_u = (x_{\bar{u}}, x_{u})$. 
	Let $\solarb_{u}$ be the subtree rooted at $x_u$, and $\solarb_{\bar{u}}$ be the subtree obtained after deleting $e_u$ and $\solarb_{u}$.
	Because $t_2$ is in $T_u$, the character $v$ cannot change along the path $P$ from $\zerom$ to $x_u$.
	So, to span $t_1$ and $t_3$, there must be edges $e_1 = (x_1, y_1)$ and $e_2 = (x_2, y_2)$ along which $v$ changes in $\solarb_{\bar{u}} \setminus P$ and $\solarb_{u}$ respectively.
	Also there must be some node $y$ along the path from $x_u$ to $x_2$ such that the subtree rooted at $y$ contains $t_2$.
	Thus if $u$ is a good character, then $v$ is a bad character.
	
	For the second part of the claim, consider the independent set formed after removing $\vc$ i.e., characters $[\dimension'] = \mset \setminus \vc$. 
	The arborescence on the terminals projected onto the independent set $[\dimension']$ can be constructed in polynomial time by \autoref{cor:pp-P-time-solvable}.
	Then, in this arborescence, the characters corresponding to $\vc$ are added back to each node (both Steiner and terminal) initially with a value $0$, and the arborescence is changed as necessary (by including relevant nodes/edges) to span $\terminalset$.
	Since every character in the minimal vertex cover conflicts with at least one character in the independent set and every character in the independent set is a good character, we can use the same proof as above to show that each of these $\vc$ characters are bad in the resulting arborescence.
\end{proof}

Due to \autoref{claim:vc-cg-relation} and as $|\bad{\optarb}| \leq \qopt$, we get the following corollary.
\begin{corollary}
	\label{cor:qopt>=|MVC|}
	The optimal penalty $\qopt \geq \tau$, where $\tau$ is the size of the minimum vertex cover of the conflict graph.
\end{corollary}

\subsubsection{Algorithm Outline}
By \autoref{claim:vc-cg-relation}, there is a \sa ~$\solarb$ such that the set of bad characters of $\solarb$ is a minimum vertex cover $\mvc$ of the conflict graph $\cg{\terminalset}$.
We present a parameterized approximation algorithm that attempts to construct such an arborescence. 
The main idea behind our proposed algorithm is to delete the characters in $\mvc$ from the input instance $I$, and thus construct the optimal \sa ~on the new instance $I'$ in polynomial time.
Finally, the removed characters are added back.
The pseudocode is given in \autoref{algo:q:approx} and is analyzed in \autoref{thm:q:approx}.

\begin{algorithm}[h]
	\caption{\label{algo:q:approx} Parameterized Approximation Algorithm for the Parameterized \msap.}
	\AlgInput{Dimension of the hypercube $\dimension$, terminal set $\terminalset$, an integer $q\geq0$.}\\
	\AlgParameter{Penalty on the cost $q$.}\\
	\AlgOutput{A near-optimal \sa~$\solarb$.}
	\begin{algorithmic}[1]
		\State Construct $\cg{\terminalset}$ and find its minimum vertex cover $\mvc$.
		\If{size of $\mvc$ $\mvcs > q$}
			\State \Return No-Instance
		\EndIf
		\State Reduce $I$ to the instance $I' = \instance{\dimension',\terminalset'}$, where $[\dimension'] = [\dimension] \setminus \mvc$ and $\terminalset' = \{t' \mid \exists~t\in \terminalset~\forall~u\in [\dimension']~t'[u]=t[u] \}$.
		\State Use \autoref{cor:pp-P-time-solvable} to compute the optimal solution $\solarb$ of $I'$.
		\State $\forall~t\in \vertexset{\solarb}$ add the $\mvc$ characters back and set them to $0$.
		\ForAll{$t \in \terminalset$}
			\State Find $t_a$ such that $t_a = \argmin\limits_{t' \in \vertexset{\solarb} \cap \anc{t}} \hdist{t',t}$.
			\State Add any path $\tpath{t_a,t}$ from $t_a$ to $t$ in $\Qm$.
		\EndFor
	\end{algorithmic}
\end{algorithm}

\qapprox*
\begin{proof}
	By \autoref{cor:qopt>=|MVC|}, if $q < \mvcs$ then $q < \qopt$; clearly this is a No-Instance.
	But when $\mvcs \leq q < \qopt$ the output of the algorithm is arbitrary.
	Henceforth, we assume $q \geq \qopt$ to prove the approximation guarantees of the output \sa.
	
	Let the vertex cover $\vc$ of $\cg{\terminalset}$ used to construct $\solarb$ be an $\alpha$-approximation of the minimum vertex cover $\mvc$.
	The skeleton of $\solarb$ will consist an optimal \sa~on $I'$, and each terminal shall be connected to its respective ancestor. 
	The structure of the output has been depicted in \autoref{fig:outputStructure}.
	\begin{figure}[H]
		\centering
		\includegraphics[scale=0.35]{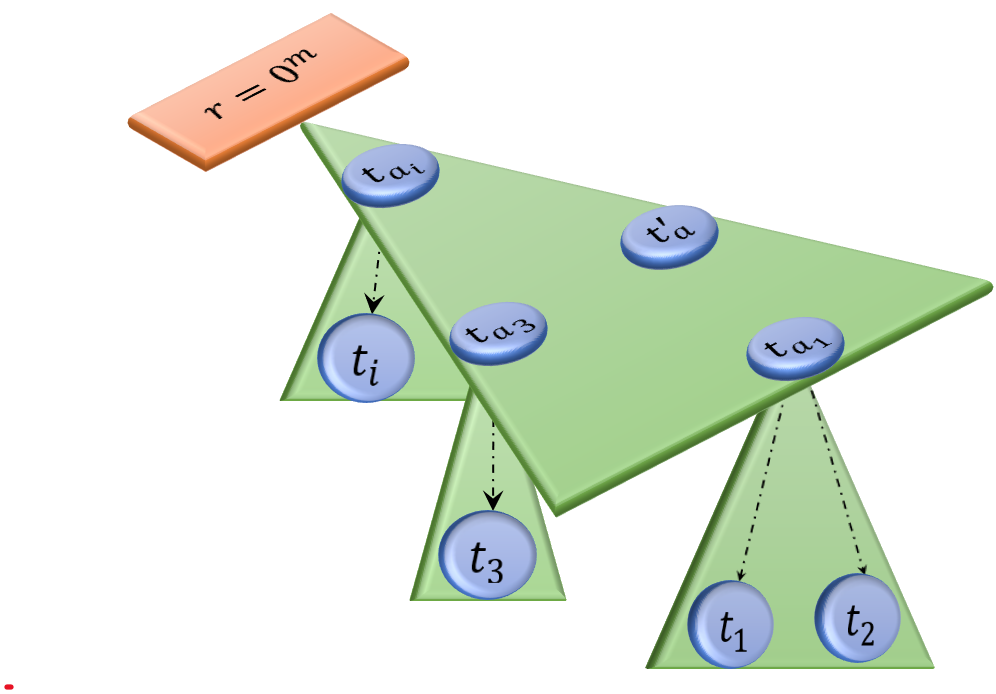}
		\caption{\label{fig:outputStructure}Structure of the output \sa~$\sol$. Triangles represent sub-trees while circles represent nodes in the arborescence. Dotted arrows represent paths.}
	\end{figure}
	
	Let the optimal \sa~of $I'$ be $\optarb^{\terminalset'}$.
	Thus, cost of the output arborescence, $\sol$, is the sum of $\cost{\optarb^{\terminalset'}}$ and the total cost of all paths to the terminals from their corresponding ancestors in $\optarb^{\terminalset'}$. 
	Now, $\cost{\optarb^{\terminalset'}} = \dimension' = \dimension-|\vc|$. 
	\begin{equation*}
		\begin{split}
			\sol &= \dimension - |\vc| + \sum_{t\in \terminalset} |\tpath{t_a,t}|
			= \dimension - |\vc| + \sum_{t\in \terminalset} \hdist{t_a,t}
			\leq \dimension - |\vc| + \sum_{t\in \terminalset} |\vc|\\
			&\leq \dimension - |\vc| + |\vc|\modr 
			= \dimension + |\vc|(\modr-1)
			\leq \opt + \alpha \mvcs \times \opt \\
			&= (1 + \alpha \mvcs) \opt \leq (1 + \alpha\qopt) \opt \leq (1 + \alpha q) \opt ~(\text{from \autoref{cor:qopt>=|MVC|} })
		\end{split}
	\end{equation*}
	If we use the \pfpt{q}-algorithm to find an exact minimum vertex cover \cite{FPTAlgoBook,FPTAlgoBook-Niedermeier,vertexCover}, then $\alpha = 1$.
	While construction of the conflict graph takes $\Oh{\modr\dimension^2}$ time, computing the minimum vertex cover can be achieved in $\Ohs{1.25284^q}$~\cite{vertexCover}.
	All the remaining parts of the algorithm can be performed in $\Oh{\modr\dimension}$ time.
	Thus the theorem holds true.
\end{proof}

\begin{remark}
	\label{remark:q:approx}
	If we use the $2$-approximation algorithm for the minimum vertex cover problem~\cite{approxAlgoBook}, then the algorithm outputs a $(1+2\qopt)$-approximation \sa~in polynomial time.
\end{remark}

\section{Parameterized Approximation Algorithm on Number of Steiner Nodes}
In this section, we present an algorithm for parameter $p$ that involves finding a minimum set of \sn ~recursively in a level that is adjacent to the subset of nodes of interest in the immediate higher level. 
We formulate computing this set as the minimum hitting set (MHS) problem.  

For some level $\ell \geq 2$, let  $\solStNode_{\ell} \subseteq \Qm^{\ell}$. 
Let $\solStNode_{\ell-1} \subseteq \Qm^{\ell-1}$ be a minimum set of nodes such that $\forall~t\in \solStNode_{\ell} ~\parent{t} \cap \solStNode_{\ell-1} \neq \phi$. 
Consider the collection of sets $\setsys = \{ \parent{t} \mid t \in \solStNode_{\ell} \}$ over the universe $\univ = \bigcup_{t\in \solStNode_{\ell}} \parent{t}$.
Clearly, $\solStNode_{\ell-1}$ is a minimum hitting set of $\setsys$, and is represented as $\solStNode_{\ell-1} = \minhitset{\setsys}$.
We abuse the notation slightly and use $ \minhitset{\solStNode_{\ell}}$ to represent $ \minhitset{\setsys}$.
Notice that, since they are at the same level in $\Qm$, every node in $\solStNode_{\ell}$ has exactly $\ell$ parents; and hence, to compute $\solStNode_{\ell-1}$, we can use the \pfpt{\ell,k} algorithm for the $\ell$-\mhsp~\cite{FPTAlgoBook-Niedermeier} ~(every set in the input set system is of size $\ell$ and $k$ is the bound on the solution size).
So, if $\terminalset_{\ell}$ is a subset of terminals at level $\ell$, we can recursively apply the \pfpt{\ell,p} algorithm for the $\ell$-\mhsp ~and construct a \sa ~on $\terminalset_{\ell}$.

Our proposed algorithm exploits this idea.
The set of all non-root terminals ($\terminalset \setminus \{\zerom\}$) are partitioned based on their levels and a \sa ~is constructed for each partition independently.
Finally, the union of all the constructed arborescences is returned.
For any partition $\terminalset_{\ell}$ of the terminal set, the arborescence is constructed using the following recurrence relation:
\begin{equation}
	\label{eqn:p:approx:recursion}
	\solStNode_i^{\ell} = 
	\begin{cases}
		\minhitset{\terminalset_{\ell}'}, & if~i=\ell-1 \\
		\minhitset{\solStNode_{i+1}^{\ell}}, & otherwise
	\end{cases}
\end{equation}
Here, $\terminalset_{\ell}' \subseteq \terminalset_{\ell}$ is the subset of terminals at level $\ell$ without any parents in the terminal set $\terminalset$ and $\solStNode_i^{\ell}$ represents the set of \sn ~at level $i$ in the \sa ~spanning $\terminalset_{\ell}'$. All the terminals in $X_{\ell} = \terminalset_{\ell} \setminus \terminalset_{\ell}'$ contain a parent in $\terminalset_{\ell-1}$, and hence are directly added as children to their respective parents.
The pseudocode is given in \autoref{algo:p:approx} and analyzed in \autoref{thm:p:approx}.

\begin{algorithm}[h]
	\caption{\label{algo:p:approx} Parameterized Approximation algorithm for the Parameterized \msap.}
	\AlgInput{Dimension of the hypercube $\dimension$, terminal set $\terminalset$, an integer $p \geq 0$.}\\
	\AlgParameter{Number of \sn~$p$}\\
	\AlgOutput{A near-optimal \sa~$\solarb$.}
	\begin{algorithmic}[1]
		\State Partition $\terminalset$ into $\terminalset_1, \terminalset_2 \ldots, \terminalset_{\maxlevel}$.
		\State Add $\terminalset_1$ as children of $\zerom$ in $\solarb$.
		\ForAll{$\ell \in [2, \maxlevel]$}
			\State Let $X_{\ell} = \{t \in \terminalset_{\ell} \mid \parent{t} \cap \terminalset_{\ell-1} \neq \phi\}$ and $\terminalset_{\ell}' = \terminalset_{\ell} \setminus X_{\ell}$.		
			\State $\forall ~t \in X_{\ell}$, add an edge $(t',t)$ to $\solarb$ for some $t' \in \parent{t} \cap \terminalset_{\ell-1}$.
			\State Add $\terminalset_{\ell}'$ to $\solarb_{\ell}$ and $\solStNode_{\ell}^{\ell} \gets \terminalset_{\ell}'$.
			\For{$i \in [\ell-1,1]$}
				\State Compute $\solStNode_i^{\ell}$ using \autoref{eqn:p:approx:recursion} with solution size at most $|\solStNode_{i+1}^{\ell}|$.
				\State $\forall~t \in \solStNode_{i+1}^{\ell}$, add an edge $(t',t)$ to $\solarb_{\ell}$ for some $t' \in \parent{t} \cap \solStNode_i^{\ell}$.
				\If{$|\solStNode_{\ell-1}^{\ell}| > p$}
					\State \Return No-Instance
				\EndIf
			\EndFor
			\State $\solarb \gets \solarb \cup \solarb_{\ell}$ (here we find the union of the edges and prune the extra nodes/edges, so that the resulting arborescence is valid).
		\EndFor
	\end{algorithmic}
\end{algorithm}

\papprox*
\begin{proof}
	We first prove that if $p \geq \popt$, then $\forall~\ell \in [2, \maxlevel]$ $|\solStNode_{\ell-1}^{\ell}| = |\minhitset{\terminalset_{\ell}'}| \leq p$.
	Let $\popt^i$ be the number of \sn~at level $i$ in $\optarb$.
	So $\sum_{i = 1}^{\maxlevel - 1} \popt^i = \popt$.
	For any level $\ell \in [2, \maxlevel]$, since every terminal $t \in \terminalset_{\ell}'$ does not have a parent in $\terminalset_{\ell-1}$, the parent of $t$ in $\optarb$ is a Steiner node.
	Also the set of \sn~spanning $\terminalset_{\ell}'$ in $\optarb$ is a hitting set.
	Hence $|\minhitset{\terminalset_{\ell}'}| \leq \popt^{\ell-1} \leq \popt \leq p$.
	Thus, if $\exists~\ell \in [2, \maxlevel]$ $|\minhitset{\terminalset_{\ell}'}| > p$, then it is a No-Instance.
	It may be that $\forall~\ell \in [2, \maxlevel]$ $|\minhitset{\terminalset_{\ell}'}| \leq p$ even when $p < \popt$ in which case the output is arbitrary.
	Henceforth we assume $p \geq \popt$ to prove the approximation guarantee of the algorithm.
	
	We first note that when the algorithm is constructing the arborescence on any level $\ell \in [2, \maxlevel]$, $\terminalset_{\ell-1}$ is already in $\solarb$.
	So every terminal $t \in X_{\ell}$ already has a parent in $\solarb$ and $\solarb_{\ell}$ is constructed only on $\terminalset_{\ell}'$.
	Also, $\forall~i \in [1, \ell-1]~ |\solStNode_i^{\ell}| \leq |\solStNode_{i+1}^{\ell}|$.
	If we are using an $\alpha$-approximation algorithm for the $\ell$-\mhsp, the total number of \sn ~in $\solarb_{\ell}$ can be computed as:
	\[p_{\solarb_{\ell}} = \sum_{i=\ell-1}^{1} |\solStNode_i^{\ell}|
	\leq \sum_{i=\ell-1}^{1} \alpha |\minhitset{\solStNode_{i+1}^{\ell}}|
	\leq \sum_{i=\ell-1}^{1} \alpha |\minhitset{\terminalset_{\ell}'}|
	\leq \sum_{i=\ell-1}^{1} \alpha \popt^{\ell-1}
	= (\ell-1) \alpha \popt^{\ell-1}
	\]
	
	Thus, the cost of the output arborescence is:
	\begin{equation}
		\begin{split}
			\sol &= \modr - 1 + \p
			\leq \opt + \sum_{\ell=2}^{\maxlevel} p_{\solarb_{\ell}}
			\leq \opt + \sum_{\ell=2}^{\maxlevel} (\ell-1) \alpha \popt^{\ell-1}\\
			&\leq \opt + (\maxlevel-1) \alpha \sum_{\ell=1}^{\maxlevel-1} \popt^{\ell}
			= \opt + (\maxlevel-1) \alpha \popt\\
			&\leq \opt + (\maxlevel-1) \alpha p
		\end{split}
	\end{equation}
	If we use the exact algorithm for the $\ell$-\mhsp~\cite{FPTAlgoBook-Niedermeier}, then $\alpha=1$ and $\sol \leq \opt + (\maxlevel-1) p$.	
	Now, the time taken to compute $\solarb_{\ell}$ is the total time to run the \pfpt{\ell,p} algorithm (depth-bounded search tree algorithm \cite{FPTAlgoBook-Niedermeier}) $\ell$ times and the time taken to add the edges i.e., $\Oh{\ell (\ell^p + |\terminalset_{\ell}'|^2 \dimension)}$.
	This is performed $\maxlevel - 1$ times, so the total time taken is $\Oh{\maxlevel^{p+2} + \maxlevel^2 \modr^2 \dimension}$.
\end{proof}

\begin{remark}
	\label{remark:p:approx}
	If we use the $\ln \modr$-approximation algorithm for the MHS problem~\cite{approxAlgoBook}, then the algorithm generates a $\Oh{\popt \maxlevel \ln \modr}$-additive approximation \sa~in polynomial time.
	Similarly, if we use the $\ell$-approximation algorithm for the $\ell$-MHS problem~\cite{approxAlgoBook-Williamson}, then the algorithm generates a $\Oh{\popt \maxlevel^2}$-additive approximation \sa.
\end{remark}

\section{Conclusion and Future Work}
We explored the \msap ~from different viewpoints of parameterization and presented \fpt ~algorithms for the same. 
Specifically, we showed that the problem is in \fpt~on the number of terminals $\modr$ and penalty $q$.
We also present a parameterized approximation algorithm for the parameter $q$ that can be improved to run in polynomial time at the expense of the approximation factor, and another parameterized approximation algorithm involving parameters $p$ and $\maxlevel$.
The major key point of all our algorithms is the polynomial dependence on $\dimension$, and hence the logarithmic dependence on the size of the underlying graph.
All the algorithms exploit the nature of the directed hypercube to accomplish this motive.

Future work could focus on the open questions of whether parameterized algorithms with better running times or approximation ratios are possible for these problems (e.g., can the exponential factor in the time complexity of the \pfpt{\modr} algorithm be improved from $3^{\modr}$ to $2^{\modr}$?). 
The \textsf{W}-hardness of the problem on the parameter $p$, remains an open question (i.e., is there an \pfpt{p} algorithm that is  independent of $\maxlevel$?). 

The Steiner arborescence problem and algorithms presented here also opens up interesting directions/applications in other fields.
\sa~may provide a data structure to represent an ensemble of related graphs (using bit vectors that indicate the presence/absence of edges in each graph), and thereby help efficiently compute certain properties of these graphs. 
An alternative representation of the terminal set as a set system may be used to explore connections to VC (Vapnik–Chervonenkis) dimension. For instance, it will be interesting to explore if the cost of the \sa~can be exploited to compute the VC dimension of a set family.

\backmatter

\bmhead{Acknowledgements}
This work was supported in part by the Wellcome Trust/DBT India Alliance Intermediate Fellowship IA/I/17/2/503323 awarded to Manikandan Narayanan.
The authors thank Dr. Vijayaragunathan Ramamoorthi for his invaluable suggestions and contribution towards the completion of this work. 

\bmhead{Author contribution}
This work is performed as part of the doctoral thesis of Sugyani Mahapatra with extensive inputs from Manikandan Narayanan and N. S. Narayanaswamy.

\bibliography{references/steiner,references/phylogeny,references/misc,references/books}

\end{document}